\documentclass[twocolumn,
nofootinbib,
prd,aps,superscriptaddress,tightenlines]{revtex4}
\usepackage{graphicx}
\usepackage{bm}
\usepackage{epsfig}

\begin{document}
\def\abs#1{ \left| #1 \right| }
\def\ket#1{ \left| #1 \right\rangle }
\def\vev#1{ \left\langle #1 \right\rangle }
\def\nn{\nonumber \\ }
\newcommand\lsim{\roughly{<}}
\newcommand\gsim{\roughly{>}}
\newcommand\CL{{\cal L}}
\newcommand\CO{{\cal O}}
\newcommand\half{\frac{1}{2}}
\newcommand\beq{\begin{eqnarray}}
\newcommand\eeq{\end{eqnarray}}
\newcommand\eqn[1]{\label{eq:#1}}
\newcommand\intg{\int\,\sqrt{-g}\,}
\newcommand\eq[1]{eq. (\ref{eq:#1})}
\newcommand\meN[1]{\langle N \vert #1 \vert N \rangle}
\newcommand\meNi[1]{\langle N_i \vert #1 \vert N_i \rangle}
\newcommand\mep[1]{\langle p \vert #1 \vert p \rangle}
\newcommand\men[1]{\langle n \vert #1 \vert n \rangle}
\newcommand\mea[1]{\langle A \vert #1 \vert A \rangle}
\newcommand\bi{\begin{itemize}}
\newcommand\ei{\end{itemize}}
\newcommand\be{\begin{equation}}
\newcommand\ee{\end{equation}}
\newcommand\bea{\begin{eqnarray}}
\newcommand\eea{\end{eqnarray}}
\def\Dsl{\,\raise.15ex \hbox{/}\mkern-12.8mu D}
\newcommand\Tr{{\rm Tr\,}}
\newcommand\lqcd{\Lambda_{\text{QCD}}}
\newcommand{\me}[3]{\ensuremath{\left\langle{#1}\vphantom{#2 #3}
\right|{#2}\left|\vphantom{#1 #2}{#3}\right\rangle}}
\newcommand{\vect}[1]{\mathbf{#1}}

\title{Enhanced Nonperturbative Effects in $Z$ Decays to Hadrons}

\author{Christian W.~Bauer}
\affiliation{California Institute of Technology, Pasadena, CA 91125}

\author{Christopher Lee}
\affiliation{California Institute of Technology, Pasadena, CA 91125}

\author{Aneesh V.~Manohar}
\affiliation{Department of Physics, University of California at San Diego,
  La Jolla, CA 92093}

\author{Mark B. Wise}
\affiliation{California Institute of Technology, Pasadena, CA 91125}

\date{February 2004}

\begin{abstract}
We use soft collinear effective field theory (SCET) to study nonperturbative strong interaction effects in $Z$ decays to hadronic final states that are enhanced in corners of phase space. These occur, for example, in the jet energy distribution for two jet events near $E_J=M_Z/2$, the thrust distribution near unity and the jet invariant mass distribution near zero. The extent to which such nonperturbative effects for different observables are related is discussed. 
\end{abstract}

\maketitle


\section{Introduction \label{sec:intro}}

Some of the most successful applications of perturbative QCD are to processes such as $Z$ decay to hadrons or  $e^+ e^-$ annihilation at large center-of-mass energy, in which a state with no strong interactions decays into final hadronic states. This paper will discuss the case of $Z$ decay, but the results in this paper apply equally well to the other cases.  Not only is the total hadronic $Z$ decay width calculable but so are less inclusive infrared-safe quantities like the $Z$ decay rate into 2-jet and 3-jet events, the thrust distribution and jet mass distributions. Comparison of perturbative predictions for these and other quantities with experimental data on $Z$ decays from LEP and SLD has led to a remarkably accurate extraction of the strong coupling constant $\alpha_s(M_Z)$~\cite{expdata1,expdata2,expdata3,expdata4,expdata5,alphasreview,compare}.
Although the extraction of the strong coupling from event shape variables is less accurate than from the total hadronic $Z$ width, it is more model-independent since (neglecting quark mass effects) it does not depend on the values of the quark couplings to the $Z$.

For the totally inclusive hadronic $Z$ decay width, the operator product expansion allows one to include in theoretical predictions nonperturbative strong interaction effects that are characterized by vacuum expectation values of local operators. The effects of higher-dimension operators are suppressed by powers of the strong interaction scale  $\lqcd$ divided by the center-of-mass energy $M_Z$.  Since the $Z$ mass is large, these effects are very small. For example, if quark masses are neglected, the leading nonperturbative effects in the $Z$ decay width come from the vacuum expectation value of the gluon field strength tensor squared, $\vev{G_{\mu \nu} G^{\mu\nu}}$. This dimension-four operator gives rise to corrections to the total hadronic width suppressed by $\lqcd^4/M_Z^4\sim10^{-9}$.

Less inclusive variables that characterize $Z$ decay to hadrons give rise to nonperturbative effects suppressed by smaller powers of $\lqcd/M_Z$~\cite{Webber,MW,KS99,KS95,KT00,DokWeb95,DokWeb97,Doktalk}. Furthermore, these corrections often become even more important in corners of phase space where hadronization effects are significant, such as in the thrust distribution very near $T=1$. It has been conjectured that the enhanced nonperturbative effects to many event shape distributions have a universal form with a single nonperturbative parameter \cite{DokWeb95,CataniWebber,KT00,Doktalk,DokMarSal,DokMarWeb}. These arguments are based on analysis of renormalon ambiguities in the QCD perturbation series and on the behavior of resummed perturbation theory. The conjectured relationship between the nonperturbative corrections to event shape distributions has recently been tested experimentally \cite{compare}. 

Recently the enhanced nonperturbative effects that occur for the jet energy distribution in corners of phase space have been studied using effective field theory methods~\cite{shape}. This approach uses the fact that very low momentum degrees of freedom which contain the nonperturbative physics couple to the degrees of freedom with energies of order $M_Z$ via Wilson lines.  Nonperturbative effects have been extensively studied previously~\cite{KS99} using factorization methods to divide the process into hard, jet-like  and soft subprocesses~\cite{factorization,StermanTasi}. Nonperturbative effects are computed from the soft subprocess. The effective field theory approach is similar to the one based on factorization methods. In this paper we elaborate on the work in~\cite{shape} and extend it to other shape variables. The enhanced nonperturbative effects are expressed in terms of weighted matrix elements of operators involving Wilson lines, where the weighting depends on the event variable being considered. We hope that this paper will help make the results of Ref.~\cite{KS99} more accessible to the community of high energy theorists who are most familiar with effective field theory methods. 

In this paper we study smeared distributions which allows us to expand the nonperturbative effects in powers of $\lqcd$, and write them as matrix elements of Wilson line operators and their derivatives. The computations are similar to those of smeared distributions in the endpoint region in $B$ decay---the point-by-point computation requires knowing the nonperturbative shape function, whereas nonperturbative effects in the smeared distributions can be written in terms of $\lambda_{1,2}$ provided the smearing region is large enough. 
 
For pedagogical reasons we start with a detailed treatment of the jet energy $E_J$ in $Z$ decay to two jets, where the jets are defined as Sterman and Weinberg did in their original work on jets in QCD~\cite{SW}. We spend considerable effort on this variable because the theoretical expression for its enhanced nonperturbative corrections is simpler than for other more phenomenologically interesting variables like thrust. Cone algorithms for jets, like that of Sterman and Weinberg, are ambiguous at higher orders in perturbation theory~\cite{Seymour,RUNIIJET}. However, since our main interest is in nonperturbative effects and we only treat the perturbative effects at order $\alpha_s$, that ambiguity is not important for us. At lowest order in perturbation theory, the $Z$ boson creates a quark and an antiquark, each with energy $M_{Z}/2$, and so the jet energy distribution is equal to
\begin{equation}
{{\rm d}\Gamma_{\text{2-jet}}\over {\rm d} E_J}=\Gamma_{\text{2-jet}}^{(0)}\ \delta(E_J-M_{Z}/2),
\end{equation}
where $\Gamma_{\text{2-jet}}^{(0)}$ is the total two-jet rate at lowest order in perturbation theory. This leading order theoretical expression for the jet energy distribution is singular at $E_J=M_Z/2$. Furthermore, the leading perturbative and nonperturbative corrections are also singular at that kinematic point.  However, a non-singular quantity that can be compared with experiment without any resummation of singular terms is obtained by smearing the jet energy distribution over a region of size $\Delta$ that contains the lowest order partonic endpoint at $E_J=M_{Z}/2$. The leading nonperturbative correction to this smeared energy distribution is suppressed by $\lqcd/\Delta$. So, for example, with $\Delta \sim 10~ {\rm GeV}$ the nonperturbative corrections are expected to be of order $10\%$, roughly the same size as perturbative corrections, and an order of magnitude larger than the order $\lqcd/M_Z$ correction expected in the complete two jet rate. We argue that for $E_J$ very near $M_Z/2$ it is not possible to capture the dominant nonperturbative effects simply by shifting, $E_J \rightarrow E_J-\mu_{\rm np}$, in the perturbative expression for ${\rm d }\Gamma_{\rm 2-jet}/{\rm d}E_J$ (where $\mu_{\rm np}$ is a nonperturbative parameter of order $\lqcd$).

In the next section, we derive an expression for the leading enhanced nonperturbative correction to the smeared jet energy distribution for two jet events using methods from soft-collinear effective field theory (SCET)~\cite{SCET1a,SCET1b,SCET2,SCET3}. This correction is given by the vacuum expectation value of a nonlocal operator involving Wilson lines. Perturbative order $\alpha_s$ corrections to this variable are derived in Appendix~\ref{sec:pert}.  

Section~\ref{sec:other} discusses the leading nonperturbative corrections for thrust, jet masses, the jet broadening variables, the $C$ parameter and energy-energy correlations. In agreement with Ref.~\cite{KS99} we find that the correction to jet mass sum and thrust are related. However, without additional model-dependent assumptions we do not find that the enhanced nonperturbative corrections to the $C$ parameter and jet broadening variables can be related to those for thrust and the jet masses. We compare the level of our understanding of the enhanced nonperturbative effects in these variables.

\section{Operator Product Expansion For The Two Jet Energy Distribution \label{sec:ope}}

The nonperturbative corrections to the energy distribution for $Z$ decay to two jets, ${\rm d}\Gamma_{\text{2-jet}}/ {\rm d} E_J$ near $E_J=M_Z/2$ are computed in this section. The perturbative corrections will be discussed in Appendix~\ref{sec:pert}. The results are given for the Sterman-Weinberg jet definition, where a cone of half-angle $\delta$ contains a jet if the energy contained in the cone is more than $E_{\text{cut}}=\beta M_Z$. We take the cone half-angle $\delta$ and the dimensionless energy cut variable $\beta$ to be of order a small parameter $\lambda$, and compute in a systematic expansion in powers of $\lambda$. We are interested in the jet energy distribution within a region $\Delta$ of $M_Z/2$, where $M_Z\gg \Delta \gg \lambda^2 M_Z$. For example,  $\Delta \sim \lambda M_Z$. The Sterman-Weinberg jet definition, like other cone algorithms, is ambiguous. However, this difficulty does not become apparent at order $\alpha_s$ in perturbation theory and also will not influence our discussion of nonperturbative effects. 

SCET is the appropriate effective field theory for the kinematic region of interest, and will be used for the derivation of the nonperturbative corrections to ${\rm d}\Gamma_{\text{2-jet}}/ {\rm d} E_J$ near $E_J=M_Z/2$. It is convenient  to introduce two lightlike vectors $n$ and $\bar n$ which satisfy $n^0=\bar n^0=1$ and ${\bf n}=-{\bf \bar n}$.   Four-vectors are decomposed along the $n$, $\bar n$ and perpendicular directions: $V=(V^+,V^-,V_{\perp})$ where $V^+ = n \cdot V$, $V^- = \bar n \cdot V$ and $ V_{\perp}^{\mu}=  V^{\mu}-V^+ {\bar n}^{\mu}/2-V^- n^{\mu}/2 $. For the problem of interest, SCET contains $n$-collinear, $\bar n $-collinear  and ultrasoft degrees of freedom~\cite{bfprs}. The $n$-collinear and $\bar n$-collinear degrees of freedom have typical momenta that scale as 
\begin{equation}
\label{moment}
p_c^{(n)} \sim M_Z(\lambda^2,1, \lambda),\qquad p_c^{(\bar n)} \sim M_Z(1,\lambda^2, \lambda),
\end{equation}
 and the ultrasoft degrees of freedom have momenta that scale as
\begin{equation}
p_{u} \sim M_Z(\lambda^2,\lambda^2,\lambda^2).
\end{equation}
 We take $\lambda \sim \sqrt{\lqcd /M_Z}$ which implies that the typical ``off-shellness" of the ultrasoft degrees of freedom, $p_{u}^2 \sim M_Z^2 \lambda ^4 \sim \lqcd^2$, is set by the QCD scale while the typical ``off-shellness" of the collinear degrees of freedom, $p_c^2 \sim M_Z^2 \lambda^2 \sim M_Z \lqcd$, is much larger than $\lqcd^2$.
Hence the collinear degrees of freedom can be treated in perturbation theory. 

The nonperturbative effects we are after are characterized by matrix elements of operators composed from the ultrasoft degrees of freedom. In $Z$ decay into two jets, the jets are almost back-to-back, and $\mathbf{n}$ is  chosen along one of the jet directions. The degrees of freedom in the two jets are then represented by $n$-collinear (for the antiquark jet) and $\bar n$-collinear fields (for the quark jet). In this section we work to lowest order in perturbation theory in the collinear fields. Hence we match the weak neutral current in full QCD onto the effective theory at tree level,
\begin{equation}
\label{current}
j^{\mu}=[\bar \xi_{\bar n}W_{\bar n}]\Gamma^{\mu}[W_n^{\dagger}\xi_n],
\end{equation}
where $\Gamma^{\mu}=g_V\gamma^{\mu}_{\perp}+g_A\gamma^{\mu}_{\perp}\gamma_5$ involves the vector and axial couplings of the $Z$ boson. The fields $\xi_{\bar n}$ and $\xi_{n}$ are collinear quark fields in the $\bar n$ and $n$ directions and we have adopted the convention
\begin{equation}
\label{sumlabels}
\xi_n(x)=\sum_{\tilde p} e^{-i \tilde{p} \cdot x}\xi_{n, \tilde p}(x),
\end{equation}
where the label momentum $\widetilde p$ contains the components of order $1$ and $\lambda$, $\bar n \cdot p$ and $\mathbf{p}_\perp$, and the order $\lambda^2$ components are associated with the space-time dependence of the fields. The Wilson lines $W_{n, \bar n}$ are required to ensure collinear gauge invariance~\cite{SCET2}. Since in this section we work to lowest order in QCD perturbation theory, they play no role in the analysis and can be set to unity.

The typical momenta of the partons in the jets are of the order of the collinear momenta, Eq.~(\ref{moment}), where the overall scale of their momentum is set by $M_Z$. However, it is possible for the jets to contain partons with momenta that have an overall scale that is much less than $M_Z$. Because of the sum over all values of $\widetilde p$ in Eq.~(\ref{sumlabels}), such partons can still be represented by collinear fields. The interaction of $n$-collinear fields among themselves is given by the full QCD Lagrangian, and so the hadronization of $n$-collinear partons into a jet is described by the full theory.

The Lagrangian of the effective theory does not contain any direct couplings between collinear particles moving in the two different lightlike directions labeled by $\bar n$ and $n$; however, they can interact via the exchange of ultrasoft gluons. It is convenient to remove the couplings of the collinear degrees of freedom to the ultrasoft ones via the field redefinition~\cite{SCET3}
\begin{equation}
\label{def1}
\xi_{n} \rightarrow Y_n^{\dagger}\xi_n,\qquad A_n \rightarrow Y_n^{\dagger} A_n Y_n,
\end{equation}
where $A_n$ is an $n$-collinear gluon field and
\begin{equation}
\label{def2}
Y_n(z)=P \exp\left[ig\int_0^\infty {\rm d}s\ n \cdot A_{u}(ns+z)\right]
\end{equation}
denotes a path-ordered Wilson line of ultrasoft gluons in the $n$ direction from $s=0$ to $s=\infty$. This is the appropriate field redefinition for outgoing collinear fields, since if a factor of $\exp(-\epsilon s)$ is inserted in the integrand to decouple the interactions at late times, one reproduces the correct $i\epsilon$ prescription for the collinear quark propagator. For annihilation which contains incoming collinear particles $Y_n$ is from $s=-\infty$ to $s=0$ and the daggers are reversed in Eq.~(\ref{def1}). An analogous field redefinition with $n \to \bar n$ removes the couplings in the Lagrangian of ultrasoft fields to the $\bar n$-collinear fields. 

\begin{widetext}
The differential decay rate for $Z$ decay to two jets is
\begin{eqnarray}
{\rm d}\Gamma_{\text{2-jet}}={1 \over 2 M_Z}\sum_{\rm {final ~states}}{1 \over 3}\sum_{\rm {\epsilon}}\abs{ \me{ J_n J_{\bar n}X_{u} } { j^{\mu}(0)\epsilon_{\mu}} { 0 } }^2 (2\pi)^4\delta^4(p_Z-p_{J_{n}}-p_{J_{\bar n}}-k_{u}),
\label{1.01}
\end{eqnarray}
where the sum over final states includes the usual phase space integrations and $\epsilon$ is the polarization vector of the decaying $Z$ boson. 
Since after the field redefinitions shown in Eq.~(\ref{def1}), there are no interactions between the ultrasoft and collinear degrees of freedom, the matrix element factorizes, and at lowest order in perturbation theory in the collinear degrees of freedom,
\begin{eqnarray}
\label{main}
{\rm d}\Gamma_{\text{2-jet}} &=& {1 \over 2 M_Z}{{\rm d}^3{\bf p}_q \over (2 \pi)^3 2 p_q^0}{{\rm d}^3{\bf p}_{\bar q} \over (2 \pi)^3 2 p_{ \bar q}^0}\abs{ {\cal M}^{(0)}_{if} }^2  \sum_{X_{u}}(2\pi)^4\delta^4(p_Z-p_{q}-p_{\bar q}-k_{u})\nn
&& \times  {1 \over N_C}\me{ 0 }{\bar T [Y_{n d}~^{e} Y^{\dagger}_{{\bar n}e}~^a](0)}{X_{u}(k_{u})}
\me{X_{u}(k_{u})}{\vphantom{\Bigr|}T [Y_{\bar n a}~^{c} Y^{\dagger}_{{ n}c}~^d](0)}{0}.
\end{eqnarray}
\end{widetext}
In Eq.~(\ref{main}), $|{\cal M}^{(0)}_{if}|^2$ is the square of the $Z \rightarrow q \bar q $ decay amplitude averaged over $Z$ polarizations and summed over the quark and antiquark spins and colors, $T $(${\bar T}$) denotes time (anti-time) ordering, $N_C$ is the number of colors, and we have explicitly displayed the color indices on the ultrasoft Wilson lines.
 
The derivation of Eq.~(\ref{main}) in many ways parallels the use of the operator product expansion to compute the deep inelastic scattering cross-section, or the rate for inclusive semileptonic $B$ decay. There is, however, one important distinction. The sum over final states in deep inelastic scattering and $B$ decay is a sum over a complete set of color singlet hadron states. In Eq.~(\ref{1.01}), one is summing over a complete set of jet and ultrasoft states. These are a complete set of partonic states, and are not necessarily color singlet states. In fact, unitarity would be violated if one separately imposed the color singlet condition on each of $\ket{J_n}$, $\ket{J_{\bar n}}$ and $\ket{X_u}$. The derivation of Eq.~(\ref{main}) is valid to the extent that the sums over partonic and hadronic states are equivalent. In jet production, the color of the fast quark that turns into a jet is eventually transferred to low energy partons during the fragmentation process. The low energy partons communicate between the different jets, and make sure the whole process is color singlet. The assumption is that this color recombination does not affect the decay rate at order $\lqcd/M_Z$.

To calculate ${\rm d}\Gamma_{\text{2-jet}}/ {\rm d} E_J$ we integrate Eq.~(\ref{main}) over the allowed values of the quark and antiquark three-momentum with the factor $\delta(E_J-p_q^0)$ inserted. This corresponds to choosing the quark jet as the ``observed'' jet. If one does not distinguish between quark and antiquark jets then Eq.~(\ref{main}) still applies since the value of ${\rm d}\Gamma_{\text{2-jet}}/ {\rm d} E_J$ when the ``observed'' jet is an antiquark jet is the same. It is convenient to work in the rest frame of the decaying $Z$, $p_Z=(M_Z,M_Z,{\bf 0}_{\perp})$, and align $\bf{\bar n}$ with the quark three-momentum ${\bf p}_{q}$.  The decomposition of the quark's four-momentum in terms of label and residual momentum, $p_q=\widetilde p_q +k_q$, has the form $p_q^+=\widetilde p_q^++k_q^+$ with $p_q^-=0$, ${\bf p}_{q \perp}=0$. (Note this means that $\widetilde {\bf p}_{q \perp}={\bf k}_{q \perp}=0$ and $k_q^-=0$.) Hence the phase space integration over quark three-momentum becomes
\begin{equation}
\int {{\rm d}^3 {\bf p}_q \over (2 \pi)^3 2 p_q^0}={1 \over 4(2\pi)^2}\sum_{\tilde p_q^+} \widetilde p_q^+ \int {\rm d} k_q^+.
\end{equation}
For the antiquark's four-momentum the decomposition into residual and label momentum is $p_{\bar q}^+=k_{\bar q}^+$, $p_{\bar q}^-={\widetilde p}_{\bar q}^-+k_{\bar q}^-$ and  ${\bf p}_{\bar q \perp}=\widetilde {\bf p}_{\bar q \perp}+{\bf k}_{\bar q \perp}$. One cannot set ${\bf p}_{\bar q \perp}=0$ by a choice of $\mathbf{n}$, since $\mathbf{n}=-\mathbf{\bar n}$, and $\mathbf{\bar n}$ has already been fixed by the direction of the quark jet.

Expressed in terms of label and residual momenta the phase space integration over antiquark three-momentum is
\begin{eqnarray}
\int {{\rm d}^3 {\bf p}_{\bar q} \over (2 \pi)^3 2 p_{\bar q}^0}&=&\sum_{\tilde p_{\bar q}}\int {{\rm d}^4 {k}_{\bar q} \over (2 \pi)^3 }\delta\bigl(({\widetilde p_{\bar q}}+k_{\bar q})^2\bigr) \nn
&=& \sum_{\tilde p_{\bar q}}\int {{\rm d}k_{\bar q}^- {\rm d^2}{\bf k}_{\bar q \perp} \over 2(2\pi)^3}{1 \over \widetilde p_{\bar q^-}}.
\end{eqnarray}
Here the delta function $\delta\bigl(({\widetilde p_{\bar q}}+k_{\bar q})^2\bigr)=\delta(\widetilde p_{\bar q}^- k_{\bar q}^+ -\widetilde {\bf p}_{\bar q \perp}^2)$ was used to do the $k_{\bar q}^+$ integration setting $k_{\bar q}^+=\widetilde {\bf p}_{\bar q \perp}^2/\widetilde p_{\bar q}^-$. 
At leading order in the SCET expansion parameter $\lambda$ the invariant matrix element ${\cal M}^{(0)}_{if}$ only depends on the label momenta $\widetilde p_q^+$ and $ \widetilde p_{\bar q}^-$. In terms of label and residual momentum the energy-momentum conserving delta function becomes
\begin{widetext}
\begin{eqnarray}
\label{delta1}
\delta^4(p_Z-p_q-p_{\bar q}-k_{u})&=&2\delta(p_Z^--p_{\bar q}^--k_{u}^-)\delta(p_Z^+-p_q^+-p_{\bar q}^+-k_{u}^+) \delta^2({\bf p}_{\bar q \perp}+{\bf k}_{u \perp})\nn
&=&2\delta_{M_Z,\tilde p_{\bar q}^-}\delta_{M_Z,\tilde p_{q}^+} \delta^2_{{\tilde {\bf p}}_{\bar q \perp},{\bf 0}}\delta(k_{\bar q}^-+k_{u}^-)\delta(k_q^+ + k_{u}^+)  \delta^2({\bf k}_{\bar q \perp}+{\bf k}_{u \perp}).
\end{eqnarray}
The relation  $k_{\bar q}^+=\widetilde {\bf p}_{\bar q \perp}^2/\widetilde p_{\bar q}^-$ and the Kronecker delta that sets $\widetilde{\bf p}_{\bar q \perp}$ to zero imply that $k_{\bar q}^+=0$, and so this variable does not appear in the penultimate delta function in Eq.~(\ref{delta1}).

Using these results gives,
\begin{eqnarray}
\label{almost}
{{\rm d} \Gamma_{\text{2-jet}} \over {\rm d } E_J}&=&{|{\cal M}^{(0)}_{if}|^2 \over 8M_Z (2\pi)}\int {\rm d}k_q^+\int {\rm d}k_{\bar q}^-{\rm d^2}{\bf k}_{ \bar q \perp}\sum_{X_{u}} \delta\left(\frac{M_Z}{2}-E_J+\frac{k_q^+}{2}\right)\delta(k_{\bar q}^-+k_{u}^-)\delta(k_q^+ + k_{u}^+) \delta^2({\bf k}_{\bar q \perp}+{\bf k}_{u \perp}) \\
&&\times {1 \over N_C}\me {0 } { \bar T [Y_{n d}~^{e} Y^{\dagger}_{{\bar n}e}~^a](0)}{X_{u}(k_{u})}
\me{ X_{u}(k_{u})}{ \vphantom{\Bigr|} T [Y_{\bar n a}~^{c} Y^{\dagger}_{{ n}c}~^d](0)}{0}
\nonumber\\
&=& {|{\cal M}^{(0)}_{if}|^2 \over 16 \pi M_Z} \sum_{X_{u}} \delta\left(\frac{M_Z}{2}-E_J-\frac{k_u^+}{2}\right) 
{1 \over N_C}\me {0 } { \bar T [Y_{n d}~^{e} Y^{\dagger}_{{\bar n}e}~^a](0)}{X_{u}(k_{u})}
\me{ X_{u}(k_{u})}{ \vphantom{\Bigr|} T [Y_{\bar n a}~^{c} Y^{\dagger}_{{ n}c}~^d](0)}{0}.\nonumber
\end{eqnarray}
We write the remaining delta function as the integral
\begin{equation}
\delta\left(\frac{M_Z}{2}-E_J-\frac{k_u^+}{2}\right)=\int \frac{{{\rm d}u }}{2\pi}\exp \left[-i\left(\frac{M_Z}{2}-E_J-\frac{k_u^+}{2}\right){u}\right]\,.
\end{equation}
\end{widetext}

At this stage the collinear degrees of freedom have been integrated out, and the matrix elements above, which involve only ultrasoft degrees of freedom, are evaluated at leading order in the SCET expansion parameter (i.e. $\lambda \rightarrow 0$). Recall that the Sterman-Weinberg jet criteria restricts particles outside the cones used to define the two jets associated with the quark and antiquark to have energy less than $E_{\text{cut}}$ which we are taking to be order $\lambda M_Z$. In the limit $\lambda \rightarrow 0$ this energy cut becomes much larger than  a typical component of an ultrasoft four-momentum. Hence, for the matrix elements of these operators, $E_{\text{cut}}$ should be taken to infinity and does not restrict these matrix elements. Similarly the cone angle is taken to be of order $\lambda$ while the typical angle between components of ultrasoft momenta is order unity. Thus the cone angle should be taken to zero in the effective theory that contains only ultrasoft degrees of freedom  and so  there is no restriction on the ultrasoft states that are summed over in Eq.~(\ref{almost}).

Using the exponential dependence on $k_{u}$ to translate the anti-time ordered product to the space-time point $u n/2$, and then using the completeness relation to perform the sum over all ultrasoft intermediate states, 
we find for the jet energy distribution
\begin{equation}
\label{doneit}
{{\rm d} \Gamma_{\text{2-jet}} \over {\rm d } E_J}=\Gamma_{\text{2-jet}}^{(0)}\ S(M_Z/2-E_J)
\end{equation}
where the shape function $S$ is defined by \cite{KS95}
\begin{eqnarray}
\label{shapefunction}
S(k)&=&{1 \over N_C}\int {{\rm d}u \over 2 \pi}{\rm e
}^{i k u}
\\
&& \times \me{ 0 }{ \bar T [Y_{n d}~^{e} Y^{\dagger}_{{\bar n}e}~^a](un/2)T [Y_{\bar n a}~^{c} Y^{\dagger}_{{ n}c}~^d](0)}{ 0},\nonumber
\end{eqnarray}
and the total two jet Z-decay width at lowest order in perturbation theory is
\begin{equation}
\Gamma_{\text{2-jet}}^{(0)}={ \abs{ {\cal M}^{(0)}_{if}}^2  \over 16 \pi M_Z }={N_C M_Z \over 12 \pi}(g_V^2+g_A^2),
\end{equation}
having implicitly summed over spins and colors.
The $n$-directed and $\bar n$-directed ultrasoft Wilson lines commute since $(s_1n-s_2\bar n)^2=-4s_1s_2<0$, and the gauge fields in the Wilson lines are space-like separated. 

In this derivation we chose the jets to be composed entirely of collinear degrees of freedom. This is appropriate since jets are confined to narrow cones. For example the momentum of any massless particle in the quark jet satisfies $p^- \ll p^+$, which is the appropriate scaling for collinear particles in the $\bar n$ direction. However, it is possible to repeat the above derivation allowing ultrasoft degrees of freedom to be inside a jet. Then instead of inserting $\delta(E_J - p_q^0)$ into Eq.~(\ref{main}), one inserts $\delta(E_J - p_q^0 - k^0_{uJ})$, where $k^0_{uJ} = (k^+_{uJ}/2)[1+{\cal O}(\lambda)]$ denotes the total ultrasoft energy inside the quark jet. Using the delta functions in Eq.~(\ref{delta1}) we obtain again Eq.~(\ref{almost}), with $k_u^+$ in the final delta function now denoting the total ultrasoft momentum \emph{outside} the quark jet. However, as mentioned previously, at leading order in $\lambda$ the cone angle of the jet shrinks to zero, and one recovers the previous result. 

It is possible to remove the time and anti-time ordering completely in the definition of the shape function $S$.  
Using the results from Appendix~\ref{A1} our expression for the shape function becomes
\begin{widetext}
\begin{eqnarray}
\label{notimeorder}
S(k)&=&{1 \over N_C}\int {{\rm d}u \over 2 \pi}{\rm e
}^{i k u}\me{ 0} { [{\overline Y^{\dagger}_n}^e~_d Y^{\dagger}_{{\bar n}e}~^a](un / 2)[Y_{\bar n a}~^{c} {\overline Y_n}^d~_c](0)}{0} \nonumber   \\
&=&{1\over N_C}\me{0}{[{\overline Y^{\dagger}_n}^e~_d Y^{\dagger}_{{\bar n}e}~^a]\delta(k+in~ \cdot \partial/2)[Y_{\bar n a}~^{c} {\overline Y_n}^d~_c]}{0}.
\end{eqnarray}
where the overline denotes an anti-triplet Wilson line.  

Since in the kinematic region of interest $M_Z/2-E_J$ is much larger than $n ~\cdot \partial$ acting on ultrasoft gauge fields it is appropriate to expand the delta function above which gives
\begin{equation}
\label{expand}
S(M_Z/2-E_J)=\delta(M_Z/2-E_J)+\delta^{\prime}(M_Z/2-E_J)\me{0}{O_1}{0} +{1 \over 2}\delta^{\prime \prime}(M_Z/2-E_J)\me{ 0}{O_2}{0}+\ldots
\end{equation}
where
\begin{equation}
\label{operators}
O_m={1\over N_C}\left[{\overline Y^{\dagger}_n}^e~_d Y^{\dagger}_{{\bar n}e}~^a\right]\left({i n~ \cdot \partial \over 2 }\right)^m\left[Y_{\bar n a}~^{c} {\overline Y_n}^d~_c\right]={1\over N_C}{\rm Tr}\,\left[Y^{\dagger}_{\bar n}\left( {in~ \cdot D \over 2}\right)^mY_{\bar n }\right].
\end{equation}
\end{widetext}
The simple form for the operators $O_m$ arises because the variable $E_J$ is totally inclusive on the ``unobserved'' antiquark jet. 

The formula for ${\rm d}\Gamma_{\text{2-jet}}/{\rm d}E_J$ is
\begin{eqnarray}
{{\rm d}\Gamma_{\text{2-jet}}\over {\rm d}E_J}&=&\Gamma^{(0)}_{\text{2-jet}}\ \Bigl[ \delta(M_Z/2-E_J) \nn
&&+\delta^{\prime}(M_Z/2-E_J)\me{0}{O_1}{0} + \ldots \Bigr]  .
\label{pert}
\end{eqnarray}
The delta function term in Eq.~(\ref{expand}) simply reproduces the leading perturbative formula for ${\rm d}\Gamma^{(0)}_{\text{2-jet}}/{\rm d}E_J$ while the higher-order terms contain the effects of nonperturbative physics. 
The derivation presented here assumes the observed jet is the quark jet. A similar derivation in the case where the antiquark jet is observed gives operators
\begin{equation}
{\overline  O_m}={1 \over N_C}{\rm Tr} \left[{\overline Y_n}^{\dagger}\left({i{\bar n}~ \cdot D \over 2}\right)^m{\overline Y_{ n }}\right].
\end{equation}
Since the vacuum expectation values of $O_m$ and ${\overline O_m}$ are equal by charge conjugation, our results also hold in the case where one does not distinguish between quark and antiquark jets.

We define the matrix elements using dimensional regularization with $\overline{{\rm MS}}$ subtraction so that in perturbation theory the vacuum expectation values $\me{ 0 }{O_m}{0 }$ are zero. 

Note that $O_2$ is a very different operator than $O_1$ so it is not possible to capture the effects of nonperturbative physics for $|E_J -M_Z/2|\sim \lambda^2 M_Z$ \footnote{More correctly the differential cross section ${\rm d}\Gamma_{\text{2-jet}}/{\rm d}E_J$ smeared over a region $\Delta$ of energy (that contains $E_J=M_Z/2$) with $\Delta$ of order $\lambda^2M_Z$.} simply by taking the lowest order perturbative formula in Eq.~(\ref{pert}) and shifting $E_J$ by a nonperturbative parameter $\mu_{\text{np}}$, i.e.  $E_J\rightarrow E_J-\mu_{\text{np}}$.  This ansatz results in the shape function
\begin{eqnarray}
S(M_Z/2-E_J)&=&\delta(M_Z/2-E_J)+\delta^{\prime}(M_Z/2-E_J)\mu_{\text{np}} \nn
&& +{1 \over 2}\delta^{\prime \prime}(M_Z/2-E_J)\mu_{\text{np}}^2+\ldots
\label{3.01}
\end{eqnarray}
where the series of derivatives of delta functions has coefficients that are simply related by $\me{ 0}{O_m}{0} =\me{ 0}{O_1}{0}^m$, which is not correct. 

For $|E_J -M_Z/2| \sim \lambda^2 M_Z$ all terms in the series of Eq.~(\ref{expand}) are equally important. However for $|E_J -M_Z/2|\sim \Delta\gg \lambda^2 M_Z$ the vacuum expectation value of $O_1$ provides the leading  order $\lqcd/\Delta$ nonperturbative correction. In this kinematic region the shift $E_J \rightarrow E_J-\mu_{\text{np}}$, with $\mu_{\text{np}}=\langle 0|O_1|0\rangle$, correctly captures the most important effects of nonperturbative physics.

We have focused on nonperturbative effects that are enhanced in the region near $E_J=M_Z/2$. If one considers a variable like the average value of the jet energy over the entire allowed phase space, then there are sources of nonperturbative corrections that we have not considered.

Using the results of Appendix~\ref{A2}, the operator $O_1$ in Eq.~(\ref{operators}) can be expressed in terms of the gluon field strength tensor \cite{KS95}:
\begin{eqnarray}
\label{fancy}
O_1&=&{1 \over 2}{\rm Tr} [Y^{\dagger}_{\bar n}(in~ \cdot D)Y_{\bar n }] \\
&=&\frac{1}{2}{\rm Tr}\left[ig\int_0^{\infty}{\rm d}s\ Y_{\bar n}^{\dagger}(z;s,0)n^{\mu}{\bar n}^{\nu}G_{\mu \nu} Y_{\bar n}(z;s,0)\right].\nonumber
\end{eqnarray}
$O_1$ in Eq. (\ref{fancy}) vanishes if the ultrasoft gauge field is treated as a classical degree of freedom. Then the Wilson lines in Eq. (\ref{fancy}) are unitary matrices and the trace vanishes since the gluon field strength tensor is in the adjoint representation. Note that  the  vacuum expectation value of $O_1$ can still be nonzero because of quantum effects. Usually operators involving products of gluon fields require renormalization, however, it is straightforward  to show that $O_1$ is not renormalized at one loop.

\section{Enhanced Nonperturbative Corrections to  Event Shape Variables \label{sec:other}}

There are a number of event shape distributions that are commonly studied in the literature.  Conventionally, one defines a general event shape distribution ${\rm d} \sigma/{\rm d}e$, where $e$ is an event shape variable defined such that the region $e \to 0$ corresponds to the two jet limit. Examples are $e=1-T$ for thrust, $e=B$ for jet broadening and $e=C$ for the $C$ parameter. Any event shape distribution in $Z$ decay contains both perturbative and nonperturbative contributions. The perturbative effects can be computed as a perturbation series in $\alpha_s(M_Z)$. At leading order, only two-jet (i.e.\ $q \bar q$) events contribute. Events with more hard partons are suppressed by powers of $\alpha_s(M_Z)$. In general, nonperturbative effects are suppressed by  powers of $\Lambda_{\rm QCD}/M_Z$, but in corners of phase space where $e \ll 1$ these nonperturbative effects become enhanced. Here we consider the region $\Lambda_{\rm QCD} \ll M_Z e \ll M_Z$ and focus on the enhanced nonperturbative contribution suppressed only by a single power of $\Lambda_{\rm QCD}/(M_Z e)$. 

Working to leading order in $\alpha_s(M_Z)$, the dominant nonperturbative effects are corrections to the two-jet distribution. Nonperturbative corrections to higher order processes are suppressed by additional powers of $\alpha_s(M_Z)$. We will compute the enhanced nonperturbative correction to some commonly measured event shape distributions, just as we did for the jet energy distribution in Sec.~\ref{sec:ope}. Recall for the jet energy distribution the dominant nonperturbative correction came from expanding 
\begin{eqnarray}
\label{EJdelta}
\delta\left(\frac{M_Z}{2}-E_J+ \frac{k_q^+}{2}\right) &=& \delta\left(\frac{M_Z}{2}-E_J\right) \\
&&+\delta^{\prime}\left(\frac{M_Z}{2}-E_J\right)\frac{k_q^+}{2}+\ldots \nonumber
\end{eqnarray}
in Eq.~(\ref{almost}) to linear order in $k_q^+$. The delta function from Eq.~(\ref{delta1}) sets $k_q^+=-k_{u}^+$, and we therefore find
\begin{equation}
\label{poor1}
{{\rm d}\Gamma^{(0)}_{\text{2-jet}}\over {\rm d}E_J}=\Gamma^{(0)}_{\text{2-jet}}\left[\delta\left(\!\frac{M_Z}{2}-E_J\!\right)\!-\!\delta^{\prime}\left(\!\frac{M_Z}{2}-E_J\!\right)\!\frac{\vev{ k_{u}^+ }}{2} \right],
\end{equation}
where
\begin{eqnarray}
\label{poor2}
\vev{ k_{u}^+ } &=& \sum_{X_{u}} {1 \over N_C}\me{ 0}{\bar T \left[Y_{n d}~^{e} Y^{\dagger}_{{\bar n}e}~^a \right] (0)}{X_{u}(k_{u})} \nn
&& \times \me{ X_{u}(k_{u})}{\vphantom{\Bigr|} T \left[Y_{\bar n a}~^{c} Y^{\dagger}_{{ n}c}~^d \right](0) }{0}  k_{u}^+.
\end{eqnarray}
The jet energy distribution has the nice property that one can write $\vev{ k_{u}^+ }$ as the vacuum expectation value of an operator involving Wilson lines of ultrasoft gauge fields [namely, Eq.~(\ref{fancy})]. For some shape variables this is not possible. However, expressions analogous to Eqs.~(\ref{poor1}--\ref{poor2})  can be derived.

\subsection{Thrust \label{ssec:thrust}}

First we consider the thrust distribution ${\rm d } \Gamma/{\rm d} T$ where the thrust $T$ is defined by
\begin{equation}
M_Z\ T=\max_{\hat \mathbf{t}} \sum_{i} \abs{\hat \mathbf{t} \cdot {\bf p _i}},
\end{equation}
where $\hat \mathbf{t}$ is a unit vector that defines the thrust axis. The maximum is taken over all possible directions of $\hat \mathbf{t}$, and the sum is over all final state particles.
To the order we are working the thrust axis $\hat \mathbf{t}$ can be set equal to the spatial part of the lightlike 
four-vector $n$ used to define the collinear antiquark field. It is convenient to call this direction the $z$-axis. The thrust distribution is calculated analogously to the two jet distribution except that the delta function $\delta(E_J-p_q^0)$  is replaced by  $\delta\left(M_ZT-\abs{p_q^z}- \abs{p_{ \bar q}^z}-\sum_\alpha \abs{k_{u\alpha}^z}\right)$, where the sum is over all ultrasoft particles. We adopt the same conventions as in the jet energy distribution so that the phase space integrals are again done using the delta function in Eq.~(\ref{delta1}). Decomposing the total ultrasoft four-momentum, $k_{u}=k_{u}^{(a)}+k_{u}^{(b)}$, into the sum of the ultrasoft momentum from particles in the same hemisphere as the antiquark (type $a$) and the same hemisphere as the quark (type $b$) the thrust $T$ can be written as
\begin{eqnarray}
\label{Tdelta}
M_ZT &=& \abs{p_q^z} +  \abs{p_{ \bar q}^z}+ \sum_\alpha \abs{k_{u\alpha}^z}\\
&=& {1 \over 2}\left(p_{\bar q}^- -p_{\bar q}^+\right)- {1 \over 2}\left(p_{q}^- -p_{q}^+\right) \nn
&&+{1 \over 2}\left(k_{u}^{(a)-}- k_{u}^{(a)+}\right) -{1 \over 2}\left(k_{u}^{(b)-}- k_{u}^{(b)+}\right)\nonumber \\
&=&{1 \over 2}\widetilde p_{\bar q}^- + {1 \over 2}\widetilde p_{q}^+ +{1 \over 2}\left(k_{\bar q}^- - k_{\bar q}^+\right)+{1 \over 2}k_q^+ \nonumber \\
&& +{1 \over 2}\left(k_{u}^{(a)-}- k_{u}^{(a)+}\right) -{1 \over 2}\left(k_{u}^{(b)-}- k_{u}^{(b)+}\right).\nonumber
\end{eqnarray}
Now the delta functions in Eq.~(\ref{delta1}) set $\widetilde p_{\bar q}^- = \widetilde p_q^+ = M_Z$, $k_{\bar q}^- = -k_u^-$, $k_q^+ = -k_u^+$, and $k_{\bar q}^+ = 0$.
Thus we find
\begin{eqnarray}
\label{thrustdef}
T &=& 1 - \frac{1}{M_Z} \left(k_{u}^{(a)+} + k_{u}^{(b)-}\right),
\end{eqnarray}
where we have also used  $k_u = k_u^{(a)} + k_u^{(b)}$. Thus,
\begin{eqnarray}
\label{thrust}
{{\rm d} \Gamma \over {\rm d} T} &=& \Gamma^{(0)}_{\text{2-jet}}\left[\delta(1-T) 
- \delta^{\prime}(1-T)\frac{\vev{ k_{u}^{(a)+} +k_{u}^{(b)-}}}{M_Z} \right]\nn
&\equiv& \Gamma^{(0)}_{\text{2-jet}}\left[\delta(1-T) 
- \delta^{\prime}(1-T)\frac{\vev{O_1^T}}{M_Z} \right].
\end{eqnarray}
The thrust axis and the hemispheres are determined by the jet directions, and can be defined in terms of the label momenta of the quark and antiquark. Thus $\hat \mathbf{t}$ and the hemispheres $a$ and $b$ are label variables. Nevertheless, because of the hemisphere condition on the ultrasoft momentum in Eq.~(\ref{thrust}), there isn't a simple formula expressing the correction in terms of the vacuum expectation value of an operator involving Wilson lines like the one in Eq.~(\ref{fancy}).  

In a region $|T-1|\sim \lambda ^2$ the higher order terms in the ultrasoft momentum that were neglected in Eq.~(\ref{thrust}) are important. Eq.~(\ref{thrust}) is appropriate for a region $\delta T$ near $T=1$  that satisfies $1 \gg \delta T \gg \lambda ^2$,  for example, $\delta T  \sim \lambda$.

\subsection{Jet Masses \label{ssec:jmass}}

The squared jet masses $M^2_{a,b}$ are the squares of the invariant mass of all the particles in the two hemispheres $a$ and $b$, defined by the plane perpendicular to the thrust axis. Two commonly used variables are the sum of jet masses, $\widehat{M}_S^2=(M_a^2+M_b^2)/M_Z^2$, and the heavy jet mass $\widehat{M}^2_H=\max(M_a^2,M_b^2)/M_Z^2$. The jet masses are $M^2_a=(p_{\bar q}+k_{u}^{(a)})^2$ and $M^2_b=(p_{ q}+k_{u}^{(b)})^2$. More explicitly,
\begin{eqnarray}
\label{MaMb}
M_a^2 &=& (p_{\bar q}^+ + k_u^{(a)+})(p_{\bar q}^- + k_u^{(a)-}) - (\vect{p}_{\bar q\perp} + \vect{k}_{u\perp}^{(a)})^2 \nn
M_b^2 &=& (p_q^+ + k_u^{(b)+})(p_q^- + k_u^{(b)-}) - (\vect{p}_{q\perp} + \vect{k}_{u\perp}^{(b)})^2.\;\;\;\;\;
\end{eqnarray}
Recall that $\vect{p}_q$ is aligned along $\bar\vect{n}$ so that $\vect{p}_{q\perp}=0$. Also, the delta function in Eq.~(\ref{delta1}) sets $\widetilde\vect{p}_{\bar q\perp} = 0$ and $\widetilde p_{\bar q}^- = \widetilde p_q^+ = M_Z$. Then, working to linear order in the ultrasoft momenta, $M^2_a = M_Z k_{u}^{(a)+}$ and $M^2_b = M_Z k_{u}^{(b)-}$, so
\begin{eqnarray}
{{\rm d} \Gamma \over {\rm d} \widehat{M}_S^2}&=&\Gamma^{(0)}_{\text{2-jet}}\left[\delta(\widehat{M}^2_S)- \delta^{\prime}(\widehat{M}_S^2)\frac{\vev{O_1^{M_S}}}{M_Z} \right],\nn
{{\rm d} \Gamma \over {\rm d} \widehat{M}_H^2}&=&\Gamma^{(0)}_{\text{2-jet}}\Bigl[\delta(\widehat{M}^2_H)
-  \delta^{\prime}(\widehat{M}_H^2)\frac{\vev{O_1^{M_H}}}{M_Z} \Bigr].
\end{eqnarray}
where
\begin{eqnarray}
\vev{O_1^{M_S}} &=& \vev{ k_{u}^{(a)+}+k_{u}^{(b)-}}\nn 
\vev{O_1^{M_H}} &=& \vev{ \max\left( k_{u}^{(a)+},k_{u}^{(b)-}\right)}
\end{eqnarray}
Note that in the kinematic region where expanding to linear order in ultrasoft and residual momentum is appropriate, the nonperturbative corrections to the $M_S^2$ and $1-T$ distributions are given by the same nonperturbative matrix element. The nonperturbative corrections to the $M_S^2$ and $M_H^2$ distributions are different.

Working to higher orders in $k_u/M_Z$, the definitions of thrust in Eq.~(\ref{thrustdef}) and of jet masses in Eq.~(\ref{MaMb}) become different beyond linear order. However, the corrections to event shape distributions at higher orders in $\Lambda_{\text{QCD}}/(M_Ze)$ come not from expanding the argument of the delta functions used to define these variables to higher orders in $k_u/M_Z$, but rather from expanding these delta functions as power series in the ultrasoft momentum, as in Eq.~(\ref{EJdelta}) for the jet energy. So even at higher orders, the enhanced nonperturbative corrections, i.e. of order $[\Lambda_{\text{QCD}}/(M_Ze)]^n$, $n>1$, come from the leading-order correction to the argument of the delta function, which are the same for thrust and jet mass sum. So the enhanced nonperturbative corrections to thrust and jet mass sum are related to all orders in $\Lambda_{\text{QCD}}/(M_Ze)$.

\subsection{Jet Broadening}

Jet broadening variables $B_{a,b}$ are  defined by
\begin{eqnarray}
B_{a,b} = {1 \over 2 M_Z} \sum_{i \in a,b} \abs{ \mathbf{p}_i \times \hat \mathbf{t} },
\label{3.16}
\end{eqnarray}
where the hemispheres $a$ and $b$ are defined as before, and $\hat \mathbf{t}$ is the thrust axis. The jet broadening variables at order $k_u/M_Z$ require knowing the thrust axis to order $k_u/M_Z$. The thrust axis $\hat\vect{t}$ maximizes $\sum_i\abs{\hat\vect{t}\cdot\vect{p}_i}$. 

\begin{figure}
\includegraphics[width=8cm]{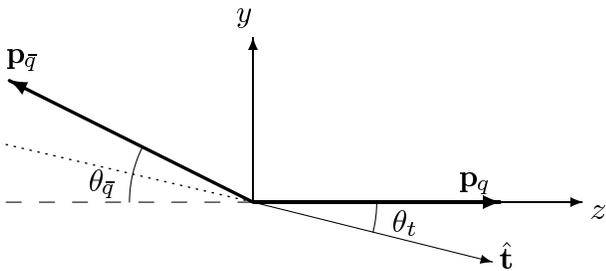}
\caption{Determination of the thrust axis. To the order we are working, the quark and antiquark have momenta $\abs{\vect{p}_q}=\abs{\vect{p}_{\bar q}}= M_Z/2$. The antiquark then makes an angle $\theta_{\bar q}= 2\abs{\vect{k}_{\bar q\perp}}/M_Z$ with the $z$-axis, and the thrust axis $\hat\vect{t}$ makes an angle $\theta_t = \abs{\vect{k}_{\bar q\perp}}/M_Z$ with both the quark and antiquark. \label{thrustaxis}}
\end{figure}
The angle between $\vect{p}_{\bar q}$ and the $z$-axis is given by 
\begin{equation}
\label{thetaqbar}
\theta_{\bar q} = \frac{\abs{\vect{k}_{\bar q\perp}}}{M_Z/2},
\end{equation}
and the thrust axis $\hat\vect{t}$ can be written as
\begin{equation}
\hat\vect{t} = (0,-\sin\theta_t,\cos\theta_t).
\end{equation}
By symmetry,
\begin{equation}
\label{thetat}
\theta_t = \frac{\abs{\vect{k}_{\bar q\perp}}}{M_Z}\,,
\end{equation}
which is half the size of $\theta_{\bar q}$ (see Fig.~\ref{thrustaxis}).

Now calculate $\abs{\vect{p}_i\times\hat\vect{t}}$ for each particle. To linear order in $k_u/M_Z$ we find for the quark,
\begin{equation}
\abs{\vect{p}_q\times\hat\vect{t}} = \frac{M_Z}{2}\sin\theta_t= \frac{\abs{\vect{k}_{\bar q\perp}}}{2},
\end{equation}
and for the antiquark,
\begin{eqnarray}
\abs{\vect{p}_{\bar q}\times\hat\vect{t}} &=& \abs{\vect{k}_{\bar q\perp}}\cos\theta_t - \frac{M_Z}{2}\sin\theta_t  = \frac{\abs{\vect{k}_{\bar q\perp}}}{2}.
\end{eqnarray}
For each ultrasoft particle $\alpha$, the cross product $\vect{k}_\alpha\times\hat\vect{t}$ is given by the determinant
\begin{equation}
\left|
\begin{array}{ccc}
\hat\vect{x} & \hat\vect{y} & \hat\vect{z} \\
k_\alpha^x & k_\alpha^y & k_\alpha^z \\
0 & -\sin\theta_t & \cos\theta_t
\end{array}
\right|.
\end{equation}
Since $\sin\theta_t$ is already of order $k_u/M_Z$ the cross product is, to linear order in $k_u/M_Z$,
\begin{equation}
\vect{k}_\alpha\times\hat\vect{t} = (k_\alpha^y,-k_\alpha^x, 0),
\end{equation}
so $\abs{\vect{k}_\alpha\times\hat\vect{t}} = \abs{{\vect{k}}_{\alpha\perp}}$.
Thus, to the order we are working, the jet broadening variables are given to this order by:
\begin{eqnarray}
B_a &=& {1 \over 2} \left[\abs{ \mathbf{k}_{\bar q\perp} } + \sum_{\alpha \in a} \abs{\mathbf{k}_{\alpha \perp} } \right] ,\nn
B_b&=& {1 \over 2} \left[\abs{ \mathbf{k}_{\bar q\perp} } + \sum_{\alpha \in b} \abs{\mathbf{k}_{\alpha \perp} } \right], 
\label{BaBb}
\end{eqnarray}
where the sum on $\alpha$ is over the ultrasoft particles in hemisphere $a$ or $b$.

One conventionally defines two other broadening variables as 
\begin{eqnarray}
\label{Bsum}
B_{\text{max}}&=&\max\left(B_a,B_b\right)\nn
B_{\text{sum}}&=&B_a+B_b
\end{eqnarray}
The jet broadening distribution is
\begin{eqnarray}
{{\rm d} \Gamma \over {\rm d}B} &=&  \Gamma^{(0)}_{\text{2-jet}} \Bigl[\delta(B) - \delta^{\prime}(B)\frac{\vev{ O_1^B }}{M_Z} \Bigr],
\end{eqnarray}
for $B_{a,b,\text{sum},\text{max}}$, where $\vev{ O_1^B } = \vev{B}$ is the matrix element of the appropriate quantity in Eqs.~(\ref{BaBb},\ref{Bsum}). Nonperturbative effects in the jet-broadening measures are not related to the jet energy or thrust.

\subsection{$C$ parameter}

The $C$ parameter is defined as
\begin{eqnarray}
C &=& 3 \left( \lambda_1 \lambda_2 + \lambda_2 \lambda_3 + \lambda_3 \lambda_1 \right),
\label{3.19}
\end{eqnarray}
where $\lambda_i$ are the eigenvalues of
\begin{eqnarray}
\theta^{rs} &=& {1 \over M_Z} \sum_i {\mathbf{p}_i^r \mathbf{p}_i^s \over \abs{\mathbf{p}_i}},
\label{3.20}
\end{eqnarray}
and $r,s=1,2,3$ are the space components of the momentum $\mathbf{p}_i$ of the $i^{\text{th}}$ particle. 

The largest component of $\theta^{rs}$ is $\theta^{zz}$. The quark and antiquark in the jets have $z$-momentum $p_q^z=-{p}_{\bar q}^z = M_Z/2$ to the order we are working. Then, to linear order in $k_u/M_Z$, the eigenvalues of $\theta^{rs}$ are given by:
\begin{eqnarray}
\label{determinant}
\det(\theta-\lambda I) = (1-\lambda) \, {\rm det} (X - \lambda I)
\end{eqnarray}
where $I$ is the identity matrix and 
\begin{eqnarray}
X_{11}&=&\sum_\alpha\frac{k_\alpha^{x2}}{M_Z\abs{\vect{k}_\alpha}}\nonumber\\
X_{22}&=&\sum_\alpha\frac{k_\alpha^{y2}}{M_Z\abs{\vect{k}_\alpha}}\nonumber\\
X_{12} = X_{21} &=& \sum_\alpha\frac{k_\alpha^x k_\alpha^y}{M_Z\abs{\vect{k}_\alpha}}
\end{eqnarray}
Here the sums over $\alpha$ are only over ultrasoft particles. (The contributions from the quark and antiquark to these components of $\theta^{rs}$ are suppressed by another factor of $1/M_Z$, since $\abs{\vect{p}_q}=\abs{\vect{p}_{\bar q}} = M_Z/2$.)

The largest eigenvalue is $\lambda_1 = 1$, and the other two eigenvalues satisfy
\begin{equation}
\lambda_2 + \lambda_3 = \frac{1}{M_Z}\sum_\alpha\frac{(k_\alpha^{x})^2 + (k_\alpha^{y})^2}{\abs{\vect{k}_\alpha}}.
\end{equation}
Thus,
\begin{equation}
C = 3  \sum_{\alpha}  { \abs{\mathbf{k}_{\alpha \perp} }^2 \over
\abs{\mathbf{k}_{\alpha} } }.
\label{3.21}
\end{equation}
The $C$ distribution is then
\begin{eqnarray}
{{\rm d} \Gamma \over {\rm d}C} &=&  \Gamma^{(0)}_{\text{2-jet}} \left[\delta(C) -\delta^{\prime}(C)
\frac{\vev{O_1^C}}{M_Z}\right].
\end{eqnarray}
where $\vev{O_1^C} = \vev{C}$ defined in Eq.~(\ref{3.21}). Like jet broadening, the $C$ parameter distribution is not local on the ultrasoft fields, and the nonperturbative correction is not related to that for any of the above distributions.

\subsection{Energy-Energy Correlation and Jet-Cone Energy Fraction}

The angular correlations of radiated energy can be characterized by the one-point and two-point correlations~\cite{BBEL,BBEL2},
\begin{eqnarray}
{ {\rm d} \Sigma \over {\rm d} \Omega} &=& \int {\rm d}\Gamma \sum_i {E_i \over M_Z} \delta\left(\Omega-\Omega_i\right) ,\nn
{ {\rm d} ^2 \Sigma \over {\rm d} \Omega{\rm d} \Omega^\prime} &=& \int {\rm d}\Gamma \sum_{i,j} {E_iE_j \over M_Z^2} \delta\left(\Omega-\Omega_i\right) \delta\left(\Omega'-\Omega_j\right), \nn
\label{5.01}
\end{eqnarray}
where the sum is over all particles, and includes the terms with $i=j$. They are normalized so that
\begin{eqnarray}
\int {\rm d}\Omega { {\rm d} \Sigma \over {\rm d} \Omega} &=&\Gamma , \nn
\int {\rm d}\Omega^\prime { {\rm d} ^2 \Sigma \over {\rm d} \Omega{\rm d} \Omega^\prime} &=& { {\rm d} \Sigma \over {\rm d} \Omega}.
\end{eqnarray}

The energy-energy correlation function $P(\cos \chi)$ is defined by
\begin{eqnarray}
P(\cos \chi) &=& \int {\rm d}\Omega^\prime {\rm d}\Omega{ {\rm d} ^2 \Sigma \over {\rm d} \Omega{\rm d} \Omega^\prime} \delta(\cos \chi - \cos \theta_{\Omega\Omega^\prime}),
\end{eqnarray}
where $\theta_{\Omega\Omega^\prime}$ is the angle between vectors in the $\Omega$ and $\Omega^\prime$ directions.

The angular energy correlations Eq.~(\ref{5.01}) were defined in Ref.~\cite{BBEL,BBEL2} for $e^+e^-$ annihilation, and the solid angle was defined with respect to the beam direction. For unpolarized $Z$ decay, there is no preferred direction, so ${\rm d}\Sigma/{\rm d}\Omega$ is a constant, and ${\rm d}^2\Sigma/{\rm d}\Omega{\rm d}\Omega^\prime$ contains the same information as the energy-energy correlation function $P(\cos\chi)$. One can, however, define distributions analogous to Eq.~(\ref{5.01}) where the solid angle is measured with respect to the thrust axis $\hat \mathbf{t}$. The one-point function is called the jet cone energy fraction $J$.

The energy-energy correlation and the jet cone energy fraction both are proportional to $\delta$ functions if one considers the leading order process of $Z$ decay into a quark-antiquark pair:
\begin{eqnarray}
P(\cos \chi) &=& J(\cos \chi) \\
&=& \frac12\Gamma_0\bigl[ \delta\left(\cos \chi - 1 \right)+ \delta\left(\cos \chi + 1 \right) \bigr].\nonumber
\end{eqnarray}
Ultrasoft emission (in two-jet events) changes the distribution in two ways: (a) by changing the energy or (b) by changing the solid angle of the emitted particles. At order $k_u/M_Z$, the change in energy can be neglected, because it does not shift the angles of the partons; thus there is no contribution proportional to $\delta^\prime(\cos \chi\pm1)$, as for variables such as thrust. The angle between the quark and antiquark is [compare Eq.~(\ref{thetaqbar})]:
\begin{equation}
\cos \theta_{q \bar q} = -1 + 2 {\mathbf{k}_\perp^2 \over M_Z^2},
\end{equation}
and the angle of the quark or antiquark with respect to the thrust axis is [compare Eq.~(\ref{thetat})]:
\begin{equation}
\cos \theta_{q \hat \mathbf{t}} = -\cos \theta_{\bar q \hat \mathbf{t}} = 1 -  {\mathbf{k}_\perp^2 \over2 M_Z^2},
\end{equation}
where $\mathbf{k}_\perp$ is the total $\perp$ momentum of the ultrasoft particles. The shift in angle is second order in $k_u/M_Z$, and so to first order, there is no enhanced contribution near $\cos \chi = \pm 1$. There are nonperturbative contributions at second order.

\subsection{Classes of Observables}

The different observables we have discussed can be divided into classes, based on the extent to which their nonperturbative corrections are inclusive on the ultrasoft degrees of freedom. 

A class I observable is the jet energy distribution. The nonperturbative correction to the jet energy depends on $\vev{k_u^+}$, where $k_u$ is the total ultrasoft momentum, so the jet energy distribution is totally inclusive on the ultrasoft fields. The derivation of nonperturbative corrections to the two jet energy distribution is not quite on the same footing as the derivation of nonperturbative corrections to the $B$ meson semileptonic decay rate, because of the additional assumption about the equivalence of sums over partonic and hadronic states discussed after Eq.~(\ref{main}).

Class II observables are thrust and the jet masses $M^2_{S,H}$. The nonperturbative corrections to these variables require the ultrasoft momentum to be broken up into two parts, $k_u=k_u^{(a)}+k_u^{(b)}$, corresponding to the contributions from ultrasoft partons in the two hemispheres. The hemispheres are chosen based on the jet directions, i.e.\ based on the collinear degrees of freedom. The momentum in each hemisphere can then be defined by integrating the ultrasoft energy-momentum tensor over the hemisphere at infinity \cite{KS99,infinity1,infinity2,infinity3,infinity4}. The class II  variables are not totally inclusive on the ultrasoft variables, but require them to be divided globally into two parts. Whether our derivation of the nonperturbative corrections for class II observables (e.g. the relation between jet mass and thrust distributions) is valid  depends on the nature of hadronization in QCD. The ultrasoft fields end up inside final state hadrons. The final hadron can contain ultrasoft partons from different hemispheres, so the hadronic energy flow in each hemisphere does not have to equal the parton energy flow in each hemisphere. If the hadronic and partonic  energy flows differ by order unity,  the derivation of nonperturbative effects in class II  observables is invalid. If, for a smearing region of size $\Delta$ the mixing of ultrasoft momenta between the two hemispheres during hadronization is an effect of order $\lqcd/\Delta$, then its impact on class II observables is the same size as $k_u^2$ effects, which are one higher order than the terms we have computed.

Class III observables are the jet broadening measures $B_{a,b,\text{sum},\text{max}}$ and the $C$ parameter. These depend on knowing the individual ultrasoft momenta of each parton. This appears to be a notion that cannot be made rigorous in field theory.

\subsection{Model-Dependent Relations among Event Shape Variables}
\label{ssec:comparison}

Nonperturbative corrections to event shape distributions have been considered extensively in the literature in the past.  For example, in the work of Ref.~\cite{KS99}, nonperturbative shape functions were derived for thrust and jet mass distributions.  The enhanced nonperturbative corrections to these distributions are given by first moments of these shape functions, and the results in sections \ref{ssec:thrust} and \ref{ssec:jmass} are in agreement with Ref.~\cite{KS99}.

The derivations of the enhanced non-perturbative corrections in this section have only relied on the fact that they arise from matrix elements of ultrasoft operators. It is insightful to understand what further conditions have to be imposed to reproduce other proposed relations amongst nonperturbative parameters for event shape distributions~\cite{DokWeb95,CataniWebber,KT00,Doktalk}. 

As an example, consider the $C$ parameter, for which the nonperturbative matrix element was defined as
\begin{eqnarray}
\langle O_1^C \rangle = 3 \left \langle \sum_\alpha \frac{|{\bf k}_{\alpha\perp}|^2}{|{\bf k}_\alpha|} \right \rangle .
\end{eqnarray}
For on-shell soft gluons collinear to the antiquark or quark jet (i.e.~in hemisphere $a$ or $b$, respectively), $k^{(a)+} \ll k^{(a)-}$ and $k^{(b)-} \ll k^{(b)+}$. This implies that
\begin{eqnarray}
\left \langle \sum_\alpha \frac{|{\bf k}_{\alpha\perp}|^2}{|{\bf k}_\alpha|} \right \rangle_{\rm coll} 
&\!\!\!\!\!\!\!=&
2 \left\langle \sum_\alpha \frac{|k^+\, k^-|}{|k^+ + k^-|} \right\rangle
\nonumber\\
&\!\!\!\!\!\!\!=&
 2 \left\langle \sum_\alpha |k^{(a)+}_\alpha| + \sum_\beta |k^{(b)-}_\beta| \right \rangle
 \nonumber\\
&\!\!\!\!\!\!\!=&
2 \left\langle k_u^{(a)}+k_u^{(b)-}\right\rangle
\end{eqnarray}
This leads to
\begin{eqnarray}
\langle O_1^C \rangle_{\rm coll} = 6 \, \langle O_1^T \rangle .
\end{eqnarray}

To take into account that ultrasoft gluons can also be radiated at a finite angle, one can impose the condition that the matrix elements of $O_1^C$ and $O_1^T$ are given by the one-gluon contribution in perturbation theory, performing the angular integrals in the phase space at a fixed value of $\abs{\vect{k}_\perp}$. Under this assumption, the matrix element of $O_1^C$ is given by:
\begin{eqnarray}
\left \langle \frac{|{\bf k}_{\alpha\perp}|^2}{|{\bf k}_\alpha|} \right \rangle _{\rm 1-gluon}
&\!\!\!\!\!\!\!=&
2\left\langle \int_0^{\frac{\pi}{2}} \!\!{\rm d}\theta\sin\theta\frac{\abs{\bf k_\perp}^2}{(\abs{\bf k_\perp}/\sin\theta)}\frac{1}{\sin^2\theta}\right\rangle
\nonumber\\
&\!\!\!\!\!\!\!=&
\pi \langle \abs{\bf k_\perp} \rangle, 
\end{eqnarray}
where the factors of $\sin\theta$ from the phase space, from the relation $\abs{\vect{k}_\perp} = \abs{\vect{k}}\sin\theta$, and from the squared amplitude for one gluon emission have all canceled out to give the final result.
For the matrix element of $O_1^T$, we calculate:
\begin{eqnarray}
\left \langle k^{(a)+} + k^{(b)-} \right \rangle_{\rm 1-gluon} 
&\!\!\!\!\!\!\!=&
2\left\langle \int_0^{\frac{\pi}{2}}\!\!{\rm d}\theta\sin\theta\frac{\abs{\bf k} (1-\cos\theta)}{\sin^2\theta} \right\rangle
\nonumber\\
&\!\!\!\!\!\!\!=&
2\left\langle \int_0^{\frac{\pi}{2}}\!\!{\rm d}\theta\frac{\abs{\bf k_\perp} (1-\cos\theta)}{\sin^2\theta} \right\rangle
\nonumber\\
&\!\!\!\!\!\!\!=& 2\langle \abs{\bf k_\perp} \rangle
\end{eqnarray}
This leads to the result
\begin{eqnarray}
\label{3pi/2}
\langle O_1^C \rangle_{\rm 1-gluon} = \frac{3 \pi}{2} \langle O_1^T \rangle_{\rm 1-gluon} .
\end{eqnarray}
Given the assumptions that have to be made to obtain Eq.~(\ref{3pi/2}) (or analogous relations based on higher orders in perturbation theory), it does not seem likely to us that there is a simple analytic nonperturbative relation  between $\vev{O_1^C}$ and $\vev{O_1^T}$.

\subsection{Comparison with the Data}

Predictions for event shape variables have been compared with  experimental data in Refs.~\cite{expdata1,compare}. Nonperturbative corrections have been included using the ansatz that their effect on distributions for shape variables is described by shifting the variable by $c \mu_{\text{np}}/E_{{\rm cm}}$ in the perturbative formula for the distribution.  Here $c$ is a constant that depends on the kinematic variable,  $\mu_{\text{np}}$ is a universal nonperturbative parameter, and $E_{{\rm cm}}$ is the center-of-mass energy. An analysis in perturbation theory (similar to what was done in section  \ref{ssec:comparison}) provides simple relations between the $c$'s for some of the event shape variables. We have found that, provided one is not in a  kinematic region that is  extremely close to the partonic endpoint (i.e. the shape function region), $c$ for $1-T$ and $M_S^2$ are the same. However, we argued that $c$ for other parameters like the heavy jet mass and $C$ are not connected to $c$ for thrust. Some experimental evidence for this can be found in the analysis of Ref.~\cite{compare}. For $1-T$ and the jet mass sum\footnote{Ref.~\cite{compare} advocates the use of a modified $E$-scheme jet mass to reduce sensitivity to hadronic masses.} a simultaneous fit for $\alpha_s$ and $\mu_{\text{np}}$ under the assumption that $c$ takes on its conjectured values (see Fig.~9 in \cite{compare}) yields values of $\mu_{\text{np}}$ that are close to each other, and values of $\alpha_s$ that are consistent with other extractions of the strong coupling. However, Ref.~\cite{compare} finds that $\mu_{\text{np}}$ for the heavy jet mass, $C$ parameter, and jet broadenings are not related to $\mu_{\text{np}}$ for thrust in the way that the analysis based on perturbation theory suggests, and, furthermore, a fit to these variables does not yield a value of $\alpha_s$ that is consistent with other extractions.

\section{Concluding Remarks \label{sec:conc}}
We have studied nonperturbative effects in $Z$ decay to hadrons using soft-collinear effective theory (SCET). The jet energy distribution for two jet events has enhanced nonperturbative effects when the jet energy is near $M_Z/2$. These nonperturbative effects can be expressed in terms of the vacuum expectation value of operators involving Wilson lines. The Wilson lines arise from the coupling of ultrasoft gluons to collinear degrees of freedom in the jet. In Appendix~\ref{sec:pert} we derive the order $\alpha_s$ perturbative corrections to
the jet energy distribution and discuss the implications of perturbative and nonperturbative physics on the first moment of this distribution.

For a region of $|E_J-M_Z/2|$ that is of size $\Delta$, the leading nonperturbative corrections to the jet energy distribution are of order $\Lambda_{{\rm QCD}}/\Delta$ when $\Delta$ is large compared to $\Lambda_{{\rm QCD}}$.  In this region they can be characterized by the vacuum expectation value of a single operator involving ultrasoft fields which provides a contribution to the jet energy spectrum that is proportional to $\delta'(M_Z/2-E_J)$. For multijet events, a similar analysis holds; however,  an additional operator analogous to $O_1$ but involving adjoint Wilson lines occurs for a gluon jet~\cite{shape}.

When $\Delta \sim \Lambda_{{\rm QCD}}$, one is in the shape function region and the functional dependence on $E_J$ is much more complicated. While we focused mostly on the kinematic region where $M_Z \gg \Delta \gg \Lambda_{{\rm QCD}}$, it was shown that in the shape function region $\Delta \sim \lqcd$, it is not possible to capture the effects of nonperturbative physics by introducing a single nonperturbative parameter $\mu_{np}$ and shifting $E_J \rightarrow E_J-\mu_{\rm np}$ in the perturbative formula for the jet energy distribution.

The jet energy distribution has the special property that it is totally inclusive in one of the jets, and hence expressions for nonperturbative effects can be derived using operator methods that are similar to those used for the endpoint region in inclusive semileptonic $B$ decay. Other event shape variables (for example, thrust, jet mass, jet broadening, etc.) have nonperturbative effects that are enhanced in the partonic endpoint region. We discussed the extent to which enhanced nonperturbative effects for these variables can be understood using field theoretic methods in QCD.

\begin{acknowledgments}

Some of this work was done at the Aspen Center for Physics. This work was supported in part by the Department of Energy under contracts  DE-FG03-97ER40546 and DE-FG03-92ER40701. C.~Lee is supported in part by a National Defense Science and Engineering Graduate Fellowship.

\end{acknowledgments}

\appendix

\begin{widetext}
\section{Properties of Wilson lines}

In this section we derive some useful properties of the ultrasoft Wilson lines introduced in Eq.~(\ref{def2}).

\subsection{Relations between triplet and anti-triplet Wilson lines}
\label{A1}

Consider the time- and anti-time-ordering of the Wilson lines in the shape function $S$ defined in Eq.~(\ref{shapefunction}). For $Y_n$, the path ordering is the same as time ordering and so $T\left[Y_n\right]=Y_n$. Consider writing $Y_n$ as the product of $N$ infinitesimal integrals over path segments of length ${\rm ds}$,
\begin{equation}
 Y_{n a}~^b= \left({\rm e}^{ig A_N ds}\right)_a~^{b_{N-1}}\left({\rm e}^{ig A_{N-1} ds}\right)_{b_{N-1}}~^{b_{N-2}}\ldots \left({\rm e}^{ig A_{1} ds}\right)_{b_{1}}~^{b},
\end{equation}
with the subscripts on the ultrasoft gauge fields denoting their space-time location along the path of integration. Taking its adjoint
\begin{equation}
Y^{\dagger}_{n a}~^b=\left({\rm e}^{-ig A_{1} ds}\right)_a~^{b_{1}}\ldots \left({\rm e}^{-ig A_{N-1} ds}\right)_{b_{N-2}}~^{b_{N-1}} \left({\rm e}^{ig A_N ds}\right)_{b_{N-1}}~^{b}.
\end{equation}
Time ordering this expression,
\begin{eqnarray}
T\left [Y^{\dagger}_{n a}~^b \right]&=& \left({\rm e}^{ig A_N ds}\right)_{b_{N-1}}~^{b}\left({\rm e}^{-ig A_{N-1} ds}\right)_{b_{N-2}}~^{b_{N-1}}\ldots \left({\rm e}^{-ig A_{1} ds}\right)_a~^{b_{1}} \nonumber \\
&&=\left({\rm e}^{-ig A^T_N ds}\right)^{b}~_{b_{N-1}}\left({\rm e}^{-ig A^T_{N-1} ds}\right)^{b_{N-1}}~_{b_{N-2}}\ldots
\left({\rm e}^{-ig A^T_{1} ds}\right)^{b_{1}}~_a={\overline Y_n}^b~_a,
\end{eqnarray}
where the overline denotes an anti-triplet Wilson line.\footnote{Recall that the generators  in the ${\overline {\bf 3}}$ representation are minus the transpose of those in the ${\bf 3}$.} Similarly,
\begin{equation}
 {\bar T} \left[Y_{{ n} a}~^b \right]={\overline Y^{\dagger}_n}^b~_a,\qquad {\bar T}\left[Y^{\dagger}_{na}~^b\right]=Y^{\dagger}_{na}~^b.
\end{equation}
From these results, Eq.~(\ref{notimeorder}) follows.

\subsection{$O_1$ in terms of the gluon field strength}
\label{A2}

We can express the operator $O_1$ in terms of the gluon field strength tensor as written in Eq.~(\ref{fancy}).
It is convenient for this purpose to generalize the expression for the ultrasoft Wilson line to
\begin{equation}
Y_{\bar n}(z;b,a)=P \exp \left[ig\int_a^b {\rm d}s\ {\bar n}\cdot A(z+\bar ns)\right]
\end{equation}
so that with $a=0$ and $b=\infty$ we recover the standard Wilson line used above, $Y_{\bar n}(z)=Y_{\bar n}(z;0,\infty)$. 
Differentiating along the $n$ direction,
\begin{eqnarray}
\label{derivative}
n \cdot \partial\ Y_{\bar n}(z) &=& ig\int_0^{\infty}{\rm ds}\ Y_{\bar n}(z;\infty,s) \left[n \cdot \partial_z \bar n\cdot
 A \right](z+{\bar n} s)Y_{\bar n}(z;s,0)\nonumber \\
&=&ig\int_0^{\infty}{\rm ds}\ Y_{\bar n}(z;\infty,s) \left[n \cdot \partial_z \bar n\cdot A-\bar n \cdot \partial _z n \cdot A+\bar n \cdot \partial _z n \cdot A \right](z+{\bar n} s)Y_{\bar n}(z;s,0) \nonumber  \\
&=&ig\int_0^{\infty}{\rm ds}\ Y_{\bar n}(z;\infty,s) \left[n \cdot \partial_z \bar n\cdot A-\bar n \cdot \partial _z n \cdot A \right](z+{\bar n} s)Y_{\bar n}(z;s,0) \nn
&&+ig\int_0^{\infty}{\rm ds}\ Y_{\bar n}(z;\infty,s) \left[{{\rm d} (n\cdot A)\over {\rm d}s}\right](z+{\bar n} s)Y_{\bar n}(z;s,0).
\end{eqnarray}
Using the chain rule,
\begin{eqnarray}
&&\int_0^\infty {\rm d}s\ {{\rm d} \over {\rm d}s} \left[Y_{\bar n}(z;\infty,s) \left[n\cdot A\right](z+{\bar n} s ) Y_{\bar n}(z;s,0)\right]= \nn
&&\quad =\int_0^{\infty}{\rm d}s\  \left[{{\rm d} \over {\rm d}s}Y_{\bar n}(z;\infty,s)\right]\left[n\cdot A\right](z+{\bar n} s ) Y_{\bar n}(z;s,0) +\int_0^\infty {\rm d}s\  Y_{\bar n}(z;\infty,s)\left[{{\rm d} (n\cdot A)\over {\rm d}s}\right](z+{\bar n} s)Y_{\bar n}(z;s,0) \nn
&&\qquad + \int_0^\infty {\rm d}s\  Y_{\bar n}(z;\infty,s)  \left[n\cdot A \right](z+{\bar n} s )\left[{{\rm d} \over {\rm d}s}Y_{\bar n}(z;s,0)\right]\nonumber  \\
&&\quad =-ig\int_0^\infty{\rm d}s\ Y_{\bar n}(z;\infty,s)[{\bar n}\cdot A(z+\bar n s),n \cdot A(z+n s)]Y_{\bar n}(z;s,0) \nn
&&\qquad+\int_0^\infty {\rm d}s\  Y_{\bar n}(z;\infty,s)\left[{{\rm d} (n\cdot A)\over {\rm d}s}\right](z+{\bar n} s)Y_{\bar n}(z;s,0) .
\end{eqnarray}
Using the above equation to eliminate the last term in Eq.~(\ref{derivative}) yields,
\begin{equation}
n\cdot D Y_{\bar n}(z)=ig\int_0^\infty {\rm d}s Y_{\bar n}(z;\infty ,s)n^{\mu} {\bar n}^{\nu} G_{\mu \nu}(z+\bar n s) Y_{\bar n}(z;s,0)
\end{equation}
where
\begin{equation}
n\cdot D Y_{\bar n}(z)=n\cdot \partial Y_{\bar n}(z)-ig n \cdot A(\infty)Y_{\bar n}(z)+igY_{\bar n}(z)n \cdot A(z)
\end{equation}
and the gluon field strength tensor is defined by,
\begin{equation}
G_{\mu \nu}= \partial_{\mu} A_{\nu}-\partial_{\nu}A_{\mu}-ig[A_{\mu},A_{\nu}].
\end{equation}
Hence
\begin{eqnarray}
O_1&=&{1 \over 2}{\rm Tr} [Y^{\dagger}_{\bar n}(in~ \cdot D)Y_{\bar n }]={1 \over 2}{\rm Tr}\left[ig\int_0^{\infty}{\rm d}s\ Y_{\bar n}^{\dagger}(z;s,0)n^{\mu}{\bar n}^{\nu}G_{\mu \nu} Y_{\bar n}(z;s,0)\right],
\end{eqnarray}
which is Eq.~(\ref{fancy}).
\end{widetext}

\section{Perturbative corrections to $d \Gamma/dE_J$ \label{sec:pert}}

\newcommand{\Diracslash}[1]{#1\hspace{-1.25ex}\slash}

Neglecting  order $\alpha_s$ corrections to the nonperturbative effects proportional to $O_1$, perturbative corrections to ${\rm d}\Gamma_{\text{2-jet}}/{\rm d}E_J$ in Eq.~(\ref{main}) can be calculated in full QCD using standard methods. In this section we first review the computation of perturbative $\mathcal{O}(\alpha_s)$ corrections to the total two-jet rate $\Gamma_{\text{2-jet}}$ and then compute the jet energy distribution ${\rm d}\Gamma_{\text{2-jet}}/{\rm d}E_J$ at order $\alpha_s$. We work in $d=4-\epsilon$ dimensions to regulate infrared, collinear and ultraviolet divergences that occur in contributions to the differential decay rate. The jets are defined using the Sterman-Weinberg criteria which involve  an energy cut $\beta M_Z$ and a cone half-angle $\delta$. Corrections suppressed by $\alpha_s \beta $ and $\alpha_s \delta$ are neglected.

\subsection{Two Jet Decay Rate \label{ssec:2jet}}

Using the Sterman-Weinberg definition of jets, there are three contributions to the two-jet rate at $\mathcal{O}(\alpha_s)$:
\begin{enumerate}
\item[(a)] One quark and one antiquark each creating a jet;
\item[(b)] One quark and one antiquark each creating a jet, plus a gluon with energy $E_g<\beta M_Z$;
\item[(c)] One quark and one antiquark each creating a jet, plus a gluon with energy $E_g>\beta M_Z$ inside one of the jets (within an angle $\delta$ of the quark or antiquark).
\end{enumerate}

Contribution (a) is simply the rate $\Gamma(Z\rightarrow q\bar q)$. The tree and virtual gluon graphs give the amplitude:
\begin{eqnarray}
\label{ampA}
\mathcal{M}_{Z\rightarrow q\bar q} = \epsilon_\mu(p_Z)\bar u^a(p_q)\Gamma^\mu v_a(p_{\bar q})\left(1 + \frac{\alpha_s C_F}{2\pi}X\right),
\end{eqnarray}
where the color index $a$ is summed over values $a=1,\dots,N_C$, and $C_F$ is the Casimir of the fundamental representation. Explicit computation of the one-loop vertex correction gives,
\begin{widetext}
\begin{eqnarray}
\label{X}
X = -\frac{4}{\epsilon^2} - \frac{3}{\epsilon} + \frac{2}{\epsilon}\ln\left(\frac{-2p_q\cdot p_{\bar q}}{\mu^2}\right) - 4 + \frac{\pi^2}{12} - \frac{1}{2}\ln^2\left(\frac{-2p_q\cdot p_{\bar q}}{\mu^2}\right) + \frac{3}{2}\ln\left(\frac{-2p_q\cdot p_{\bar q}}{\mu^2}\right).
\end{eqnarray}
Integrating the square of the amplitude over the $d$ dimensional two body phase space gives:
\begin{eqnarray}
\label{Zqq}
\Gamma_{Z\rightarrow q\bar q} &=& \frac{N_C}{32\pi^2}(g_V^2+g_A^2)\left[M_Z^{1-\epsilon}(4\pi)^\epsilon\frac{2-\epsilon}{3-\epsilon}\Omega_{3-\epsilon}\right] \\
&&\times\Biggl[1 + \frac{\alpha_s C_F}{\pi}\Bigl(-\frac{4}{\epsilon^2} - \frac{3}{\epsilon} + \frac{2}{\epsilon}\ln\frac{M_Z^2}{\mu^2} - 4 + \frac{7\pi^2}{12}  - \frac{1}{2}\ln^2\frac{M_Z^2}{\mu^2} + \frac{3}{2}\ln\frac{M_Z^2}{\mu^2}\Bigr)\Biggr],\nonumber
\end{eqnarray}
where $\Omega_{d}$ is the total solid angle in $d$ dimensions. The $1/\epsilon$ poles will cancel out against divergences from the real gluon emission graphs. We do not need to expand the bracketed prefactor in Eq.~(\ref{Zqq}) in powers of $\epsilon$ because the identical factor will appear in the real gluon graphs.

Contributions (b) and (c) come from integrating the square of the amplitude for real gluon emission, $Z\rightarrow q\bar q g$, over the three-body phase space in $d$ dimensions. We find for the terms that do not vanish as $\beta$ and $\delta$ go to zero,
\begin{eqnarray}
\Gamma_{Z\rightarrow q\bar q g}^{(b)} = \frac{g_s^2 M_Z^{1-\epsilon}N_C C_F}{256\pi^5}(2\pi)^{2\epsilon}\Bigl(\frac{\mu}{M_Z}\Bigr)^\epsilon\Omega_{2-\epsilon}\Omega_{3-\epsilon}(g_V^2 + g_A^2)
\Bigl(-\frac{1}{\epsilon}\Bigr)\frac{\Gamma(-\frac{\epsilon}{2})^2}{\Gamma(-\epsilon)}\beta^{-\epsilon},
\end{eqnarray}
and
\begin{eqnarray}
\Gamma^{(c)}_{Z\rightarrow q\bar q g} &=& \frac{g_s^2 M_Z^{1-\epsilon}N_C C_F}{256\pi^5}(4\pi)^{2\epsilon}\Bigl(\frac{\mu}{M_Z}\Bigr)^\epsilon\Omega_{2-\epsilon}\Omega_{3-\epsilon}\frac{2-\epsilon}{3-\epsilon}(g_V^2+g_A^2)\frac{(2\delta )^{-\epsilon}}{-\epsilon}\biggl[\frac{4}{\epsilon}(2\beta)^{-\epsilon} + 2\Bigl(1+\frac{3\epsilon}{4}+\frac{13\epsilon^2}{8}\Bigr)\frac{\Gamma(-\epsilon)^2}{\Gamma(-2\epsilon)}\biggr].\nn
\end{eqnarray}
Adding together contributions (b) and (c), expanding in powers of $\epsilon$ and converting to the $\overline{\text{MS}}$ scheme yields for the total rate for $Z\rightarrow q\bar q g$ in the two-jet region
\begin{eqnarray}
\label{Zqqg}
\Gamma_{Z\rightarrow q\bar q g} &=& \frac{N_C C_F\alpha_s}{32\pi^3}(g_V^2+g_A^2)\left[M_Z^{1-\epsilon}(4\pi)^\epsilon\frac{2-\epsilon}{3-\epsilon}\Omega_{3-\epsilon}\right] \\
&&\times\left[\frac{4}{\epsilon^2} + \frac{3}{\epsilon} - \frac{2}{\epsilon}\ln\frac{M_Z^2}{\mu^2}  - \frac{3}{2}\ln\frac{M_Z^2}{\mu^2} + \frac{1}{2}\ln^2\frac{M_Z^2}{\mu^2} - 4\ln 2\beta\ln\delta - 3\ln \delta + \frac{13}{2} - \frac{11\pi^2}{12}\right].\nonumber
\end{eqnarray}
Finally, we  add together the rates $\Gamma_{Z\rightarrow q\bar q}$ and $\Gamma_{Z\rightarrow q\bar q g}$ from Eqs.~(\ref{Zqq}) and (\ref{Zqqg}). The $\epsilon$-dependent prefactors in brackets in the two equations are identical, as promised. The $1/\epsilon$-poles in the remainder of the expressions cancel out exactly (as do all the logarithms of $M_Z/\mu$), so we can set $\epsilon=0$ in the remaining finite parts, leaving
\begin{eqnarray}
\label{SW}
\Gamma_{\text{2-jet}} = \frac{N_C M_Z}{12\pi}(g_V^2+g_A^2)\Biggl[1
+ \frac{\alpha_s C_F}{\pi}\Bigl(\frac{5}{2} - \frac{\pi^2}{3}- 3\ln\delta - 4\ln 2\beta\ln\delta \Bigr)\Biggr],
\end{eqnarray}
which agrees with Sterman and Weinberg's original result~\cite{SW}.

\subsection{Differential Decay Rate \protect{${\rm d}\Gamma_{\text{2-jet}}/{\rm d} E_J$} \label{ssec:diff}}

We now turn our attention to the differential decay rate ${\rm d}\Gamma_{\text{2-jet}}/{\rm d} E_J$. The contribution of $\Gamma_{Z\rightarrow q\bar q}$ to this rate is simply
\begin{equation}
\frac{{\rm d}\Gamma_{Z\rightarrow q\bar q}}{{\rm d}E_J} = \Gamma_{Z\rightarrow q\bar q}\ \delta\left(E_J-\frac{M_Z}{2}\right),
\end{equation}
where $\Gamma_{Z\rightarrow q\bar q}$ is the total rate for $Z\rightarrow q\bar q$ calculated to $\mathcal{O}(\alpha_s)$, which is given by Eq.~(\ref{Zqq}).

For the contribution of real gluon emission processes, we write the three-body phase space for this rate:
\begin{eqnarray}
\frac{{\rm d}\Gamma_{\text{2-jet}}}{{\rm d}E_J} &=& \frac{1}{16M_Z}\frac{1}{(2\pi)^{2d-3}}\Omega_{d-2}\Omega_{d-1}{\rm d}E_1 E_1^{d-4} {\rm d}E_2 E_2^{d-4} {\rm d}\cos\theta\sin^{d-4}\theta \\
&&\times\delta\biggl[\frac{M_Z^2-2M_Z(E_1+E_2)}{2E_1 E_2} + 1-\cos\theta\biggr]\delta(E_J - \cdots)|\mathcal{M}|^2,\nonumber
\label{3body}
\end{eqnarray}
\end{widetext}
where the $\delta(E_J - \cdots)$ defines $E_J$ according to which partons actually go inside the jet. It is useful to split up the phase space slightly differently than for the case of the total rate:
\begin{enumerate}
\item[(a)] Gluon with energy $E_g>\beta M_Z$ inside unobserved jet;
\item[(b)] Gluon with any energy inside observed jet;
\item[(c)] Gluon with energy $E_g<\beta M_Z$ outside observed jet.
\end{enumerate}
These three regions exhaust the possible gluon energies and locations with respect to the jets. It is convenient to introduce the variable
\begin{equation}
e_J= {M_Z \over 2}-E_J
\end{equation}
and focus on a region of $e_J$ near the origin with size of order $\beta M_Z$.

For case (a), where a gluon with $E_g>\beta M_Z$ is inside the unobserved jet, take $E_1 = E_g$, $E_2 = E_{\bar q}$, so $\theta$ is the angle between the gluon and antiquark, and $E_J=E_q$. Integrating over $\theta$ and $E_{\bar q}$ using the delta functions leaves an integral over $E_g$ running between the limits
\begin{equation}
E_g^\pm = \frac{M_Z}{4}\biggl(1\pm\sqrt{1-\frac{8e_J}{M_Z\delta^2}}\biggr),
\end{equation}
and restricts $e_J$ to lie between
\begin{equation}
\delta^2\beta M_Z<e_J<\frac{M_Z\delta^2}{8}.
\end{equation}

Similarly, for case (b), where a gluon with any energy lies inside the observed jet, $E_1 = E_q$, $E_2 = E_g$, and $E_J=E_g +E_q$. Integrate over $\theta$ and $E_q$ using the delta functions. Then the limits of the $E_g$ integral are
\begin{equation}
E_g^\pm = \frac{M_Z}{4}\biggl(1\pm\sqrt{1+\frac{8e_J}{M_Z\delta^2}}\biggr),
\end{equation}
and $e_J$ is restricted to the region
\begin{equation}
-\frac{M_Z\delta^2}{8}<e_J<0.
\end{equation}

Before we proceed to case (c), note that the physical observable we actually want to calculate is the smeared distribution
\begin{equation}
\frac{{\rm d}\Gamma}{{\rm d}E_J}\biggr|_{\Delta} = \int {\rm d}E_J w_\Delta(E_J)\frac{{\rm d}\Gamma}{{\rm d}E_J},
\end{equation}
where $w_\Delta$ is a smooth function which smears the differential rate over a region of jet energy whose size is of order $\beta M_Z$. But the contributions to the rate from cases (a) and (b) have support only over a region of size $\delta^2 M_Z\ll \beta M_Z$ near $e_J = 0$. Consider smearing ${{\rm d}\Gamma}/{{\rm d}E_J}$, or equivalently ${{\rm d}\Gamma}/{{\rm d}e_J}$, over a region near $E_J = M_Z/2$ ($e_J=0$) of size of order $\beta M_Z$. Then $w(0)\sim 1/\beta M_Z$, $w'(0)\sim 1/(\beta M_Z)^2$, etc.\ Expanding,
\begin{equation}
\int {\rm d}e_J w(e_J)\frac{{\rm d}\Gamma}{{\rm d}e_J} = \int {\rm d}e_J[w(0)+w'(0)e_J+\cdots]\frac{{\rm d}\Gamma}{{\rm d}e_J}.
\end{equation}
Since $w'(0)/w(0)\sim 1/\beta M_Z$, and, for the contributions in cases (a) and (b), $e_J \sim\delta^2 M_Z$ in the region where ${{\rm d}\Gamma}/{{\rm d}e_J}$ is nonzero, the second term is suppressed by a power of $\delta^2/\beta\ll 1$. Thus only the first term is relevant.\footnote{This argument assumes that the integral $\int {\rm {\rm d}}e_J\,e_J {\rm d}\Gamma/{\rm d}e_J$ is finite, which can easily be shown.} Keeping only the first term amounts to replacing the full ${{\rm d}\Gamma}/{{\rm d}e_J}$ by
\begin{equation}
\frac{{\rm d}\Gamma}{{\rm d}e_J}\rightarrow\delta (e_J)\int {\rm d}e_J^\prime \frac{{\rm d}\Gamma}{{\rm d}e_J^\prime}.
\end{equation}
However, integrating the contributions of (a) and (b) to ${{\rm d}\Gamma}/{{\rm d}e_J}$ over all allowed values of $e_J$ simply gives their contribution to the total Sterman-Weinberg jet rate, that is, they will build up part of the term $\Gamma_{Z\rightarrow q\bar q g}\delta(E_J-M_Z/2)$ in ${\rm d}\Gamma_{\text{2-jet}}/{\rm d}E_J$. Since we have already calculated the total rate, we need not analyze cases (a) and (b) any further, as long as we can get the remaining contribution to the total rate from case (c).

In case (c) we have a gluon with $E_g<\beta M$ anywhere outside the observed jet. Here $E_1=E_q$, $E_2=E_g$, and  $E_J = E_q$. Writing out the formula for the rate explicitly,
\begin{eqnarray}
\frac{{\rm d}\Gamma^{(c)}}{{\rm d}E_J} &=& \frac{1}{16M_Z}\frac{1}{(2\pi)^{2d-3}}\Omega_{d-2}\Omega_{d-1}\\
&&\times \theta(e_J)\theta[\beta M_Z(1-\delta^2)-e_J] \nonumber \\
&&\times\int_{e_J (1+\delta^2)}^{\beta M_Z}\!\!{\rm d}E_g\,E_g^{d-4}E_q^{d-4}\sin^{d-4}\theta|\mathcal{M}^{(c)}_{Z\rightarrow q\bar q g}|^2. \nonumber
\end{eqnarray}
The part of the amplitude that gives a contribution that survives as $\beta\rightarrow 0$  is
\begin{equation}
\label{ampB}
\abs{ \mathcal{M}^{(c)}_{Z\rightarrow q\bar q g} }^2 \!\! = 4N_C C_F g_s^2\mu^\epsilon\frac{d-2}{d-1}(g_V^2+g_A^2)\frac{M_Z^2 p_q\cdot p_{\bar q}}{(k\cdot p_q)(k\cdot p_{\bar q})}.
\end{equation}
Substituting  Eq.~(\ref{ampB}) into the phase space 
\begin{eqnarray}
\label{plus}
\frac{{\rm d}\Gamma^{(c)}}{{\rm d}E_J} &=& \frac{M_Z g_s^2 N_C C_F}{256\pi^5}\Bigl(\frac{\mu}{M_Z}\Bigr)^\epsilon(2\pi)^{2\epsilon}\frac{d-2}{d-1}\\
&&\times\Omega_{d-2}\Omega_{d-1}(g_V^2+g_A^2)e_J^{-1-\frac{\epsilon}{2}} \theta(e_J)\nonumber \\
&&\times\theta[\beta M_Z(1-\delta^2) - e_J]\ln\left[\frac{\beta M_Z - e_J}{e_J\delta^2}\right].\nonumber
\end{eqnarray}
The factor, $(1/e_J)\ln [(\beta M_Z - e_J)/(e_J\delta^2)]$,
is singular as $e_J\to 0$, and must be rewritten in terms of an integrable quantity. Use the ``plus distribution'':
\begin{equation}
\int_0^{\beta M_Z} \!\!\!\!{\rm d}e_J \,f(e_J)_+\ g(e_J) \equiv \int_0^{\beta M_Z} \!\!\!\!{\rm d}e_J\,f(e_J)[g(e_J)-g(0)],
\end{equation}
where $f$ diverges at $e_J=0$ and $g$ is a test function finite at $e_J =0$. To replace $f$ by $f_+$, we would write
\begin{eqnarray}
\int_0^{\beta M_Z} \!\!\!\!{\rm d}e_J\,f(e_J)g(e_J) &=& \int_0^{\beta M_Z}\!\!\!\! {\rm d}e_J\,f(e_J)_+\ g(e_J) \qquad\\
&&\quad + g(0)\int_0^{\beta M_Z}\!\!\!\! {\rm d}e_J\,f(e_J).\nonumber
\end{eqnarray}
The second term amounts to replacing
\begin{equation}
f(e_J)\rightarrow \delta(e_J)\int {\rm d}e_J'\,f(e_J').
\end{equation}
But making this replacement in Eq. (\ref{plus}) means writing a delta function $\delta(E_J-M_Z/2)$ and integrating the differential rate over all allowed values of $e_J$, which again just gives its contribution to the total Sterman-Weinberg jet rate. Together with the contributions from (a) and (b) this gives the one loop contribution to $\delta(E_J-M_Z/2)\Gamma_{\rm 2-jet}$. Only the plus function piece gives a deviation of the jet energy distribution away from $E_J=M_Z/2$. The final result for the differential rate to $\mathcal{O}(\alpha_s)$ is:
\begin{eqnarray}
\frac{{\rm d}\Gamma_{\text{2-jet}}}{{\rm d}E_J} &=& \delta\Bigl(E_J - \frac{M_Z}{2}\Bigr)\Gamma_{\text{2-jet}} \label{59}\\ 
&&+\frac{M_Z\alpha_s N_C C_F}{12\pi^2}(g_V^2+g_A^2)\nn
&&\times \theta(e_J)\theta(\beta M_Z-e_J)\biggl[\frac{1}{e_J}\ln \Bigl(\frac{\beta M_Z - e_J}{e_J\delta^2}\Bigr)\biggr]_+, \nonumber
\end{eqnarray}
where the total rate $\Gamma_{\text{2-jet}}$ is given by Eq.~(\ref{SW}).

\subsection{First Moment of the Jet Energy Distribution}

As an application of the above  result consider the first moment of the jet
energy distribution, defined by
\begin{equation}
M_1(f) =\int_{{M_Z \over 2}-f \beta M_Z}^{ { M_Z }} 
\hspace{-2em}{\rm d}
E_J \left( {1 \over\Gamma_{{\rm 2-jet}}} {{\rm d} \Gamma_{{\rm 2-jet}}\over
{\rm d} E_J}\right)\left( {1\over 2}-{E_J\over M_Z} \right),
\end{equation}
Using the expression in Eq.~(\ref{pert}) for the nonperturbative correction and in Eq.~(\ref{59}) for the order $\alpha_s$
perturbative correction to the jet energy distribution gives
\begin{eqnarray}
M_1(f)&=&{\alpha_s  C_F \beta  \over \pi}\left[f \log \left({1 \over f
\delta^2}\right)-(1-f)\log(1-f)\right] \nonumber \\
&&\quad+\frac{\langle 0|O_1|0\rangle}{M_Z},
\end{eqnarray}
for $f<1$ and
\begin{equation}
M_1(f)={\alpha_s C_F \beta  \over \pi}\log \left({1 \over
\delta^2}\right)+\frac{\langle 0|O_1|0\rangle}{M_Z},
\end{equation}
for $f>1$.
Note that the order $\alpha_s$ contribution to  $M_1(f)$ is independent of $f$ for $f>1$. This occurs because the perturbative correction vanishes for $E_J <M_Z/2-\beta M_Z$. 

\begin{figure}
\includegraphics[width=9cm]{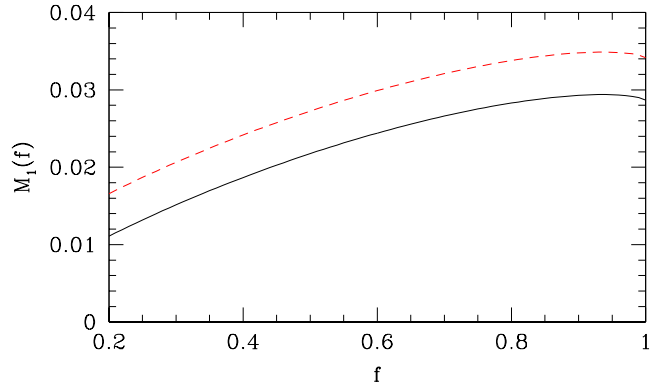}
\caption{Plot of the function $M_1(f)$. The black solid curve shows the perturbative contributions only, while the red dashed line represents the moment including the nonperturbative contribution. The figure corresponds to $\beta=0.15$, $\delta=\pi/12$, $\langle 0|O_1|0\rangle = 0.5~{\rm GeV}$, and we have evaluated the strong coupling constant at the scale $\mu = \beta M_Z$.\label{momentfig}}
\end{figure}
In Fig.~\ref{momentfig}  we plot $M_1(f)$, for $f<1$. For this figure the value of the
energy cut is $\beta=0.15$ and the cone half-angle is $\delta=15^\circ$ and the vacuum expectation value of $O_1$ is set equal to $500~{\rm MeV}$. We evaluate $\alpha_s$ at the scale $\beta M_Z$ and find with
these parameters that the order $\alpha_s$ corrections reduce the two jet
rate by about $16\%$ from its tree level value lending support to the validity of perturbation theory for the values of the cone angle and energy cut used in Fig.~\ref{momentfig}.

\subsection{Perturbative Corrections in the Effective Theory \label{ssec:match}}

Although we have used full QCD to calculate the jet energy distribution it is possible to do the computation in the effective theory. Here we briefly discuss how that computation would proceed.

The full theory amplitude for $Z\rightarrow q\bar q$ is reproduced in SCET by the Wilson coefficient in the current matching:
\begin{equation}
j_{\rm QCD}^\mu = [\bar\xi_{\bar n} W_{\bar n}]C(\mu,\widetilde p_q\cdot \widetilde p_{\bar q})\Gamma^\mu[W_n^\dag\xi_{ n}].
\end{equation}
where there is an implicit sum over label momenta  and the matching coefficient $C(\mu,\widetilde p_q\cdot \widetilde p_{\bar q})$ can be read off\footnote{The matching coefficient is just given by the finite part of the full theory matrix element $\langle q\bar q | j^\mu | 0\rangle$ because the full theory current has no anomalous dimension, so the $1/\epsilon$ poles are pure IR divergences, which must cancel out in the matching condition. The loop graphs in the effective theory contributing to this matrix element are zero in dimensional regularization, so the finite part of the matching coefficient is just the finite part of the QCD matrix element, given by Eqs.~(\ref{ampA}) and (\ref{X}), while the infinite parts become the UV counterterm in the effective theory~\cite{EFT,EFTnotes}.} from  Eqs.~(\ref{ampA}) and (\ref{X}):
\begin{eqnarray}
C(\mu,\widetilde p_q\!\cdot\!\widetilde p_{\bar q}) &=& 1 + \frac{\alpha_s C_F}{2\pi}\!\Biggl[-4 + \frac{\pi^2}{12} - \frac{1}{2}\ln^2\!\left(\!\frac{-2\widetilde p_q\cdot \widetilde p_{\bar q}}{\mu^2}\!\right) \nn
&&+ \frac{3}{2}\ln\left(\frac{-2\widetilde p_q\cdot \widetilde p_{\bar q}}{\mu^2}\right)\Biggr],
\end{eqnarray}
and the UV renormalization factor for the current in the effective theory is
\begin{equation}
Z_V = 1 + \frac{\alpha_s C_F}{2\pi}\left[-\frac{4}{\epsilon^2} - \frac{3}{\epsilon} + \frac{2}{\epsilon}\ln\left(\frac{-2\widetilde p_q\cdot \widetilde p_{\bar q}}{\mu^2}\right)\right].
\end{equation}
Note that both the renormalization factor and the matching coefficient depend on the label momenta for the quark and antiquark.
For outgoing particles, the collinear Wilson lines are  defined as
\begin{equation}
W_n(z) = P\,\exp\biggl[ig\int_0^\infty ds\,\bar n\cdot A_n(\bar n s + z)\biggr],\end{equation}
and one  must include collinear gluons produced by a Wilson line in real gluon emission to get the correct $Z \rightarrow q \bar q g$ amplitude. We find that the perturbative expressions for the two jet rate presented in the previous sections are reproduced by the effective theory if we call any particles inside the observed quark jet $\bar n$-collinear particles and all other particles $ n$-collinear. In the effective theory, ultrasoft gluons in the final state contribute zero in perturbation theory and appear only in the nonperturbative shape function. A similar result holds for deep inelastic scattering~\cite{dis}.


\end{document}